\documentclass[aps, pra, 10pt, twocolumn, tightenlines, letterpaper, amsmath, amssymb, preprintnumbers, floatfix, longbibliography, nofootinbib,dvipsnames, superscriptaddress]{revtex4-2}

\usepackage{dsfont}
\usepackage{amsmath}
\usepackage{amssymb}
\usepackage{extarrows}
\usepackage{physics}
\usepackage{tabu}
\usepackage{tabularx}
\usepackage{bm}
\usepackage{tikz}
\usetikzlibrary{shapes.geometric}
\usepackage{graphicx}
\usepackage{placeins}
\usepackage{textgreek}
\usepackage{multirow}
\usepackage{makecell}
\usepackage{soul}
\usepackage{diagbox}
\usepackage{xcolor}
\usepackage[normalem]{ulem}
\usepackage[export]{adjustbox}
\usepackage{float}
\usepackage{wasysym}
\usepackage{setspace}
\usepackage{booktabs}
\usepackage{physics}
\usepackage{quantikz}
\usepackage{enumerate}

\usepackage[hypertexnames=false]{hyperref}
\hypersetup{
    colorlinks=true,       
    linkcolor=blue,          
    citecolor=blue,        
    filecolor=blue,      
    urlcolor=blue           
}
\usepackage[capitalize]{cleveref}
\newcolumntype{Y}{>{\centering\arraybackslash}X}

\usepackage{orcidlink}

\definecolor{violet_sp}{RGB}{148,130,157}
\definecolor{blue_sp}{RGB}{165,168,234}
\definecolor{orange_trot}{RGB}{244,158,134}
\definecolor{brown_trot}{RGB}{211,149,139}
\definecolor{yellow_em}{RGB}{239,160,68}

\renewcommand{\thetable}{\arabic{table}}
\begin{document}

\begin{figure}
  \vskip -1.cm
  \leftline{\includegraphics[width=0.15\textwidth]{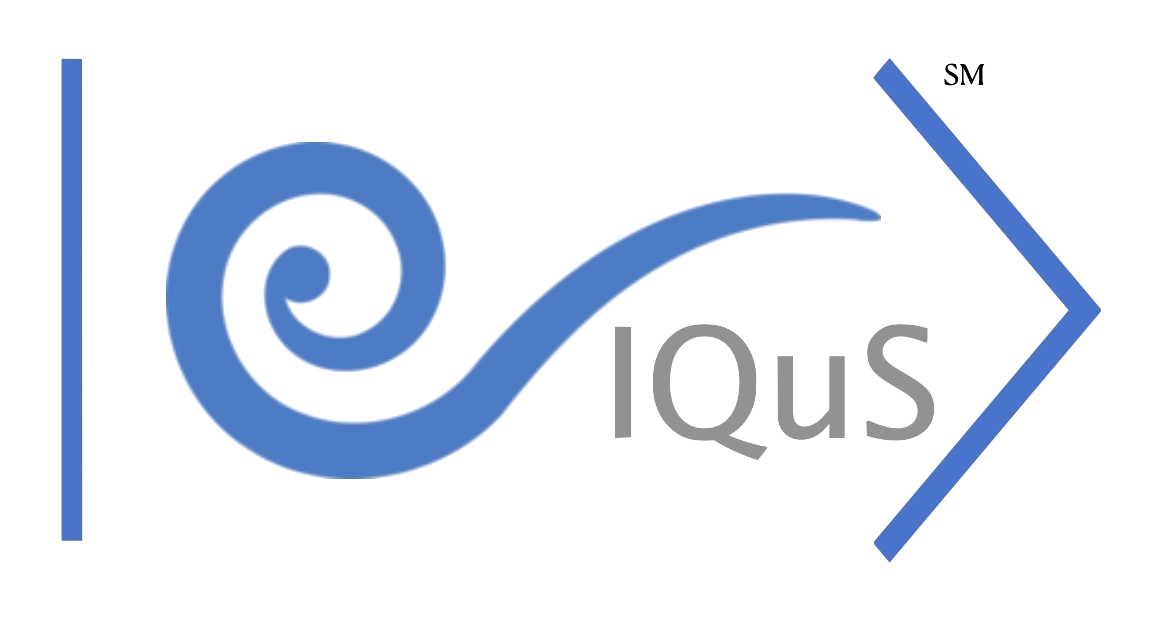}}
  \vskip -1.cm
\end{figure}

\title{Pathfinding Quantum Simulations of Neutrinoless Double-$\beta$ Decay}

\author{Ivan A.~Chernyshev
}
\affiliation{Theoretical Division, Los Alamos National Laboratory, Los Alamos, NM 87545, USA}

\author{Roland C.~Farrell
}
\affiliation{Institute for Quantum Information and Matter (IQIM) and Department of Physics, California Institute of Technology, CA 91125, USA}

\author{Marc Illa
}
\affiliation{InQubator for Quantum Simulation (IQuS), Department of Physics, University of Washington, Seattle, WA 98195, USA}

\author{Martin J.~Savage
}
\email{mjs5@uw.edu}
\affiliation{InQubator for Quantum Simulation (IQuS), Department of Physics, University of Washington, Seattle, WA 98195, USA}

\author{Andrii Maksymov
}
\affiliation{IonQ Inc., 4505 Campus Dr, College Park, MD 20740, USA}

\author{Felix Tripier
}
\affiliation{IonQ Inc., 4505 Campus Dr, College Park, MD 20740, USA}

\author{Miguel Angel Lopez-Ruiz
}
\affiliation{IonQ Inc., 4505 Campus Dr, College Park, MD 20740, USA}

\author{Andrew Arrasmith
}
\affiliation{IonQ Inc., 4505 Campus Dr, College Park, MD 20740, USA}

\author{Yvette de Sereville
}
\affiliation{IonQ Inc., 4505 Campus Dr, College Park, MD 20740, USA}

\author{Aharon Brodutch
}
\affiliation{IonQ Inc., 4505 Campus Dr, College Park, MD 20740, USA}

\author{Claudio Girotto
}
\affiliation{IonQ Inc., 4505 Campus Dr, College Park, MD 20740, USA}

\author{Ananth Kaushik
}
\affiliation{IonQ Inc., 4505 Campus Dr, College Park, MD 20740, USA}

\author{Martin Roetteler
}
\email{martin.roetteler@ionq.co}
\affiliation{IonQ Inc., 4505 Campus Dr, College Park, MD 20740, USA}

\date{\today}

\begin{abstract}
\noindent
We present results from co-designed quantum simulations of the neutrinoless double-$\beta$ decay of a 
simple 
nucleus in 1+1D quantum chromodynamics using IonQ's Forte-generation trapped-ion quantum computers.
Electrons, neutrinos, and up and down quarks
are distributed across two lattice sites and mapped to 32 qubits, with an additional 4 qubits used for flag-based error mitigation. 
A four-fermion interaction is used to implement
weak interactions, 
and lepton-number violation is induced by a neutrino Majorana mass.
Quantum circuits that prepare the initial nucleus and time evolve with the Hamiltonian containing the strong and weak interactions are executed on IonQ Forte Enterprise.
Enabled by tuned model parameters, 
lepton-number violation is observed in real time, providing
a clear signal of neutrinoless double-$\beta$ decay.
This was made possible by co-designing the simulation
to  maximally utilize the all-to-all connectivity and native gate-set available on IonQ’s quantum computers. 
Quantum circuit compilation techniques and co-designed error-mitigation methods, informed from executing benchmarking circuits with up to 2,356 two-qubit gates, enabled observables to be extracted with high precision.
We 
discuss the potential of future quantum simulations to provide yocto-second resolution of 
the reaction pathways in these, and other,
nuclear processes.
\end{abstract}

\maketitle

\section{Introduction}
\label{sec:intro}
\noindent
Future  quantum computers are 
expected to enable ab initio investigations of key unsolved problems in nuclear physics (NP) and high-energy physics (HEP) research~\cite{Dean:2018,Cloet:2019wre,Banuls:2019bmf,Klco:2021lap,Bauer:2022hpo,Catterall:2022wjq,Humble:2022klb,Beck:2023xhh,Bauer:2023qgm,DiMeglio:2023nsa}.
A central focus in the current era of noisy intermediate-scale quantum (NISQ)~\cite{Preskill:2018jim} computers is the co-design of efficient quantum algorithms, circuit compilers and error mitigation schemes to progress toward addressing specific  scientific objectives.
These advances aim to maximize the utility of quantum computers that feature imperfect gates acting on a limited number of qubits (or qudits).
Such NISQ devices are still powerful tools for simulating out-of-equilibrium dynamics, e.g., Refs.~\cite{Dumitrescu:2018njn,Klco:2018kyo,Lu:2018pjk,Klco:2019evd,Baroni:2021xtl,Ciavarella:2021nmj,Ciavarella:2021lel,Lee:2023urk,Turro:2023dhg,Zhang:2023hzr,Farrell:2024fit,De:2024smi,Zhu:2024dvz,Gonzalez-Cuadra:2024xul,Cochran:2024rwe,Zemlevskiy:2024vxt,Hayata:2024fnh,Farrell:2025nkx,Alexandrou:2025vaj,Schuhmacher:2025ehh,Davoudi:2025rdv,Chai:2025qhf}, and can provide valuable insight into mechanisms underlying quantum many-body phenomena.

An intriguing future application of quantum computers is the simulation of nuclear reactions in real-time, providing temporal snapshots of nuclei during decays, fission, fusion and more. 
These processes involve characteristic time-scales separated by many orders of magnitude.
Such multi-scale problems present significant challenges, even for quantum computers.
In this work, we propose using quantum computers to image nuclear dynamics on the shortest of these time-scales;
a yocto-second ($10^{-24}$ seconds 
$\equiv 1~{\rm ys}$).
We do not consider high-energy probes, such as beyond-TeV scale collisions, in this discussion.
This is the time scale relevant to hadronic structure with, e.g., the $\Delta$-resonance decaying to a proton with a half-life $\tau_{\Delta}\sim 5~\rm {ys}$.
The analogous 
(experimental)
development of femto-second ($10^{-15}$ seconds) imaging  in the 1990s~\cite{Zewail_femto} gave chemists access to the intermediate states that molecules pass through during chemical reactions, and revealed how atoms re-arrange during the breaking and formation of chemical bonds.
The probing of sub-yocto-second dynamics 
using quantum simulations
would provide analogous insight(s) into nuclear processes. 
Snapshots of the quantum state of nuclei on these extremely short time scales are, in principle, accessible via Hamiltonian evolution on a quantum computer.

In this work, we investigate a potential exotic nuclear decay relevant to 
searches for new physics.
The stability of matter places tight constraints on 
the structure of physics beyond the Standard Model~\cite{Weinberg:1967tq,Glashow:1961tr,Salam:1968rm,Gross:1973id,Politzer:1973fx,Higgs:1964pj},
including upper bounds on the amount that fundamental symmetries can be broken obtained from proton-decay, neutron-antineutron-oscillations, and the $\beta$-decay and $\beta\beta$-decay of nuclei, see, e.g., Refs.~\cite{Gonzalez-Alonso:2018omy,Cirigliano:2019wao,ParticleDataGroup:2024cfk,PhysRevD.110.112011,particles8010006,Pritychenko_2025}.
In the case at hand, the neutrinoless double $\beta$-decay ($0\nu\beta\beta$-decay) of certain nuclei can only occur if one of the (accidental) symmetries of the Standard Model (lepton number) is broken.
Determining the rates of such decays is a forefront
theoretical and computational challenge
due to the Majorana-neutrino~\cite{Majorana:1937} induced
process, requiring two charged-current weak interactions connected by a near-massless neutrino propagating across the nucleus (see, e.g., Refs.~\cite{Haxton:1984ggj,Elliott_2015}).
Progress toward robust computations of decay rates of nuclei continues to be 
impressive~\cite{Agostini:2022zub,Sevestrean:2024hxk,Horoi:2024wgs,PhysRevLett.124.232501,Castillo:2024jfj,PhysRevLett.132.182502,Cirigliano:2025vye},
including with Euclidean-space lattice quantum chromodynamics (QCD) simulations, e.g., Refs.~\cite{Tiburzi:2017iux,Shanahan:2017bgi,Cirigliano:2020yhp,Nicholson:2018mwc,Tuo:2019bue,Davoudi:2020xdv,Davoudi:2020gxs,Detmold:2022jwu,Davoudi:2024ukx},
but 
keeping track of the coherent sum over low-energy excitations of the nucleus during this process, along with the strong correlations between nucleons, is a task that may be better suited for 
quantum, rather than classical, computation.

Future simulations aided by quantum computers could help resolve two puzzles in the Standard Model: the nature of the neutrino mass and the mechanism behind the matter/anti-matter asymmetry in our universe.
This is because a lepton number violating neutrino mass would be intimately tied to the 
matter/anti-matter imbalance created during the electroweak phase transition in the early universe.
These fundamental questions about nature have motivated an internationally-coordinated experimental program~\cite{EXO-200:2019rkq,GERDA:2020xhi,CUORE:2021mvw,KamLAND-Zen:2022tow,
CUPID:2022puj,Majorana:2022udl,KamLAND-Zen:2024eml,AMoRE:2024loj,
LEGEND:2025jwu} searching for the $0\nu\beta\beta$-decay of nuclei,
and synergistic theoretical and computational efforts~\cite{Adams:2022jwx,Agostini:2022zub,Barabash,Lewitowicz:2025qlr}.
Importantly, an experimental observation of $0\nu\beta\beta$-decay would provide unambiguous 
evidence for the violation of lepton number.
A more detailed discussion on the potential scientific insight gained by, 
and challenges associated with, simulating $0\nu\beta\beta$ is given in 
Supplementary Note 1.

The purpose of the present work is to establish the current capabilities of trapped-ion quantum computers to simulate a rare process of current experimental and theoretical focus that has the potential to reveal new insights into fundamental physics.
Our work is a step along a path that is expected to lead to quantum 
simulations that can be used to determine or constrain new physics from corresponding experimental results.
We leverage  the power of co-designed simulations to observe
lepton-number violating dynamics on a quantum computer for the first time.
This is accomplished via the simulation of the $0\nu\beta\beta$-decay of a simple nucleus in 1+1D lattice QCD using IonQ's Forte-generation quantum computers.
Specifically, we perform lattice simulations of the decay of two baryons restricted to two spatial sites.
Both strong and weak interactions are included, and a Majorana neutrino mass term explicitly violates lepton-number conservation.
The coupling constants and masses are deliberately tuned to recover a mass hierarchy that kinematically favors double-$\beta$ decay, but suppresses single-$\beta$ decay (in this volume).

Our work builds off previous quantum simulations of non-abelian gauge theories in 1+1D dimension~\cite{Farrell:2022wyt,Farrell:2022vyh,Than:2024zaj,Klco:2019evd,Atas:2022dqm,Atas:2021ext} and beyond~\cite{Ciavarella:2024fzw,Turro:2024pxu,Ciavarella:2023mfc,Ciavarella:2021lel,Ciavarella:2021nmj,Kavaki:2024ijd,ARahman:2022tkr,ARahman:2021ktn,Balaji:2025afl,Chawdhry:2023jks}, with a particular emphasis on Ref.~\cite{Farrell:2022vyh}, where single $\beta$-decay was simulated using a similar setup.
Our quantum circuits are designed to maximally benefit from the all-to-all connectivity and native gate-set available on IonQ's  trapped-ion quantum computers.
Additionally, we introduce  techniques for mitigating statistical and device errors that are tailored to be maximally effective for the observables we measure.
These co-designed elements enable high-fidelity real-time simulations of doubly-weak decays
on a two spatial site (32 qubit) lattice.
A $10\sigma$ signal for the dynamical generation of lepton-number violation mediated by a Majorana neutrino is obtained from running circuits with 470 two-qubit gates on IonQ Forte Enterprise.
This work 
establishes
a potential path forward 
for future quantum simulations that would impact searches for new physics.

\section{Results}
\subsection{\texorpdfstring{$0\nu\beta\beta$-decay in 1+1D QCD}{}}
\noindent
A model for the $0\nu\beta\beta$-decay of a nucleus is simulated in
1+1D lattice QCD with dynamical quarks (up and down) and leptons (electrons and neutrinos).
A lattice with periodic boundary conditions (PBCs) and $L=2$ spatial lattice sites is mapped to 32 qubits of IonQ's Forte-generation quantum computers. 
A minimum of two spatial sites is needed to support the degrees of freedom produced in $0\nu\beta\beta$-decay.
The weak interactions are modeled with an effective four-Fermi interaction that locally couples quarks to leptons~\cite{Farrell:2022vyh}.
The hadronic states of 1+1D QCD form isospin multiplets~\cite{Farrell:2022wyt}, with the lowest-lying baryon multiplet having $I=3/2$, containing $\Delta^{++}, \Delta^+, \Delta^0, \Delta^-$ labeled after its similarity with the $\Delta$ resonance in 3+1D QCD (the superscripts denote the electric charge).
Parameters in the Hamiltonian, including a Majorana mass term, are tuned to permit the $0\nu\beta\beta$-decay of a $|\Delta^- \Delta^-\rangle$ initial state.

\begin{figure}[t]
    \centering
    \includegraphics[width=\linewidth]{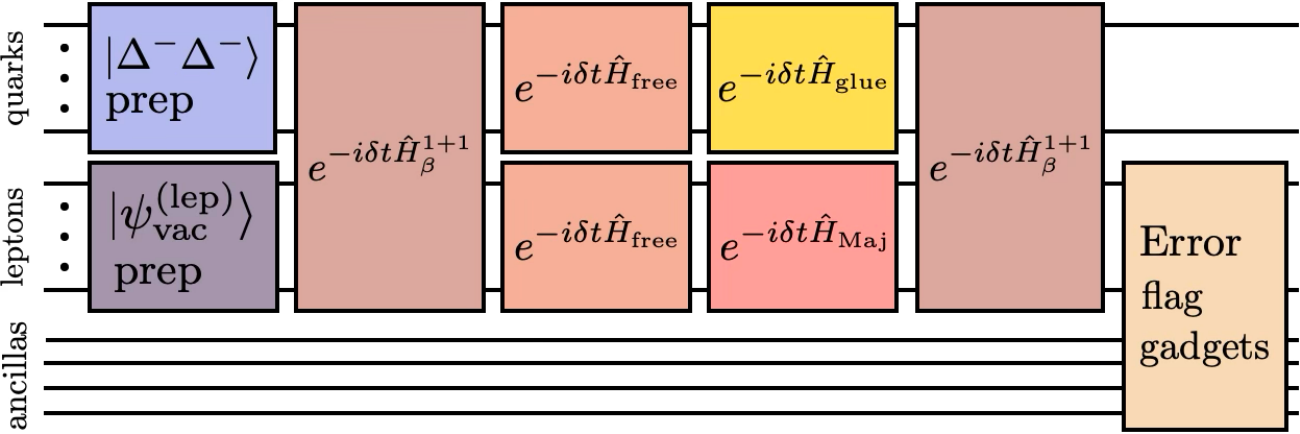}
    \caption{The structure of the quantum circuits used to simulate $0\nu\beta\beta$-decay. 
    First, the initial $|\Delta^- \Delta^- \rangle$ state is prepared using SC-ADAPT-VQE~\cite{Farrell:2023fgd}.
    Next, weak decay dynamics are implemented using two steps of Trotterized time evolution with time step $\delta t=t/2$.
    The first Trotter step has been simplified because $|\Delta^- \Delta^- \rangle$ is a strong interaction eigenstate.
    Before the final measurement, leakage events are flagged by coupling the lepton qubits to a register of ancillas. 
    All qubits are initialized in $|0\rangle$ and measured in the z-basis.
    Decompositions of each circuit block are provided in Supplementary Notes 10 and 5.
    }
  \label{fig:full_circuit_diagram}
\end{figure}

To simulate this decay, a quantum circuit that initializes the lepton vacuum and $|\Delta^- \Delta^-\rangle$ is applied, and then time evolved for time $t$ with two Trotter steps of a Hamiltonian containing the strong and weak interactions, as well as the free fermion terms.
The inclusion of a lepton-number breaking neutrino Majorana mass term in the Hamiltonian $\hat{H}_{\rm Maj}$  opens the $0\nu\beta\beta$ decay channel.

To reduce the number of two-qubit gates, we implement several approximations in the time evolution.
The chromoelectric interaction is truncated
beyond $\lambda=1$ staggered sites
and we only keep
the terms in the four-Fermi weak interaction, $\hat{H}_{\beta}^{1+1}$, that act on valence fermions.
Additionally, two-qubit rotations with angles $\theta \leq t/32$ are removed.
The effects of these approximations are detailed in Supplementary Note 2. 
Despite the errors due to approximation becoming significant for $t\geq 1.0$, our simulations are still able to extract qualitatively correct signals of $0\nu\beta\beta$-decay.
After using IonQ's circuit compiler, the required circuits have 470 $R_{ZZ}$ gates.
A schematic of the quantum circuit(s) used in our  simulations is shown in Fig.~\ref{fig:full_circuit_diagram}.

Weak decays can be detected and classified
by measuring the total electric charge of the electrons $\hat{Q}_e$ and the lepton number $\hat{{\cal L}}$.
These observables are
\begin{align}
{\hat Q}_e & \ = \  - \frac{1}{2}\sum_{n=0}^3 \hat{Z}_{25+2n} \ ,  \nonumber \\ 
\hat{{\cal L}} & \ = \ \frac{1}{2}\sum_{n=0}^{3}\left (\hat{Z}_{24+2n} + \hat{Z}_{25+2n} \right ) \ .
\label{eq:qe_lnum}
\end{align}
Both the lepton electric charge and lepton number are zero in the initial state, and a non-zero lepton electric charge signals a decay. 
Further, deviations of the lepton number from zero are due to the neutrino Majorana mass.
While not independent from these two quantities, the neutrino number $\hat{N}_\nu = \hat{Q}_e + \hat{\mathcal{L}} $ 
can be helpful in revealing
the contributions from neutrinoless decays when the Majorana mass is non-zero, $m_M\neq 0$.

\begin{figure*}[!ht]
\centering
\includegraphics[width=\textwidth]{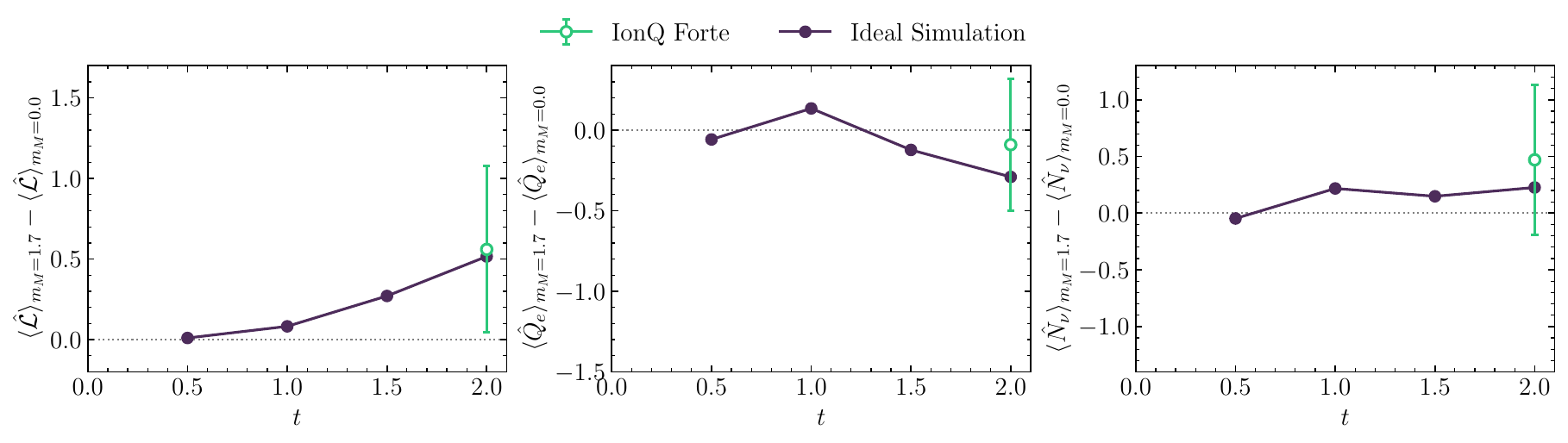}
\caption{The time evolution of the difference of the lepton number (left), 
lepton electric charge (center) and neutrino number (right) during the decay of 
the $|\Delta^-\Delta^-\rangle$ two-baryon state
in 1+1D QCD. 
Results are obtained from two steps of the first-order Trotterized circuit (2,356 native two-qubit gates) executed on IonQ Forte, and are derived from those given in Table~\ref{tab:2400_gate_data}, displayed in green, as well as noiseless statevector simulator 
(Ideal Simulation), displayed in black. 
Both sea- and valence-weak interactions are included.
The error bars are obtained from bootstrap resampling and represent one standard deviation. The gray dotted line is added for reference.}
\label{fig:F_results}
\end{figure*}
\begin{table*}[!ht]
\begin{ruledtabular}
\begin{tabular}{cccccccc}
 \multicolumn{4}{c}{Ideal Simulation} & \multicolumn{4}{c}{QPU Results} \\ \\[-1em]
 \cmidrule(lr){1-4} \cmidrule(lr){5-8}
 \\[-1em]
 $\langle\hat{\mathcal{L}}\rangle_{m_M=0}$ & $\langle\hat{\mathcal{L}}\rangle_{m_M=1.7}$& $\langle\hat{Q}_e\rangle_{m_M=0}$ & $\langle\hat{Q}_e\rangle_{m_M=1.7}$ &  $\langle\hat{\mathcal{L}}\rangle_{m_M=0}$ & $\langle\hat{\mathcal{L}}\rangle_{m_M=1.7}$& $\langle\hat{Q}_e\rangle_{m_M=0}$ & $\langle\hat{Q}_e\rangle_{m_M=1.7}$
 \\\\[-1em]
 \midrule \\[-1em]
  0.0 & 0.57 & -0.67 & -0.76 & $-0.02 \pm 0.36$ & $0.54 \pm 0.37$ & $-0.13 \pm 0.29$ & $-0.22 \pm 0.29$ \\ \\[-1em]
 \end{tabular}
\end{ruledtabular}
\caption{Observables at $t=2$ in the decay of the $|\Delta^- \Delta^-\rangle$ initial state obtained from a noiseless statevector simulator 
(Ideal Simulation) and from
executing quantum circuits with 2,356 native two-qubit gates on IonQ Forte (QPU results).
The uncertainties are estimated from bootstrap resampling 
and represent one standard deviation.
Differences between these results are shown in Fig.~\ref{fig:F_results}.
}
\label{tab:2400_gate_data}
\end{table*}
%

\subsection{Estimating the Limits of IonQ Forte}
\label{sec:big_circuit}
\noindent
In the spirit of co-design and benchmarking, we present results from simulations of $0\nu\beta\beta$-decay that pushed against the outer limits of cutting-edge quantum computers.
Specifically,
we probed the boundaries of IonQ Forte's capabilities by executing circuits with $\sim 5\times$ more two-qubit gates.
These experiments simulated $0\nu\beta\beta$-decay with fewer approximations applied to the time-evolution operator.
The full four-Fermi weak interaction, including both valence and sea components, given in Supplementary Note 3A,  
was implemented and small rotation angles were retained.
The required quantum circuits had 2,356 two-qubit gates compared to the 470 two-qubit gates executed on Forte Enterprise.
A total of 24 twirled variants (see Methods' subsection 'Circuit-Optimization and Error-Mitigation'), each with 420 shots, were run, leading to a total of 10,080 shots.

Results obtained from Forte
are shown
in Fig.~\ref{fig:F_results} for $t=2.0$, and given in Table~\ref{tab:2400_gate_data}. 
Compared to the experiments run on Forte Enterprise 
in Results~\ref{sec:IonQ_Enterprise}, 
the uncertainties from Forte
are significantly larger due to the increased noise in deeper circuits. 
The uncertainties estimated from bootstrap resampling are known to be (very) conservative 
as they include elements of the hardware-bias cancellation (while the mean values are close, the uncertainty from bootstrapping is much larger still than that from the un-bootstrapped result).
This leads to a deviation of the bootstrapped mean from the true mean, and illustrates that,
in this instance, bootstrap resampling 
provides biased estimators.

From these simulations, we concluded that implementing the full weak operator was not practical with the current generation of hardware.
The quantum circuits were too deep for error mitigation to be applied successfully.
Consequently, we reduced the circuit depth by truncating
the weak operator to only include valence fermion operators 
for subsequent simulations.  
These runs also informed the error mitigation strategy utilized on IonQ Forte Enterprise as discussed in~\ref{sec:Forte_codesign}.

\subsection{Observation of \texorpdfstring{$0\nu\beta\beta$-decay in 1+1D QCD}{} using IonQ Forte Enterprise}
\label{sec:IonQ_Enterprise}
\begin{figure*}[!ht]
\centering
\includegraphics[width=\textwidth]{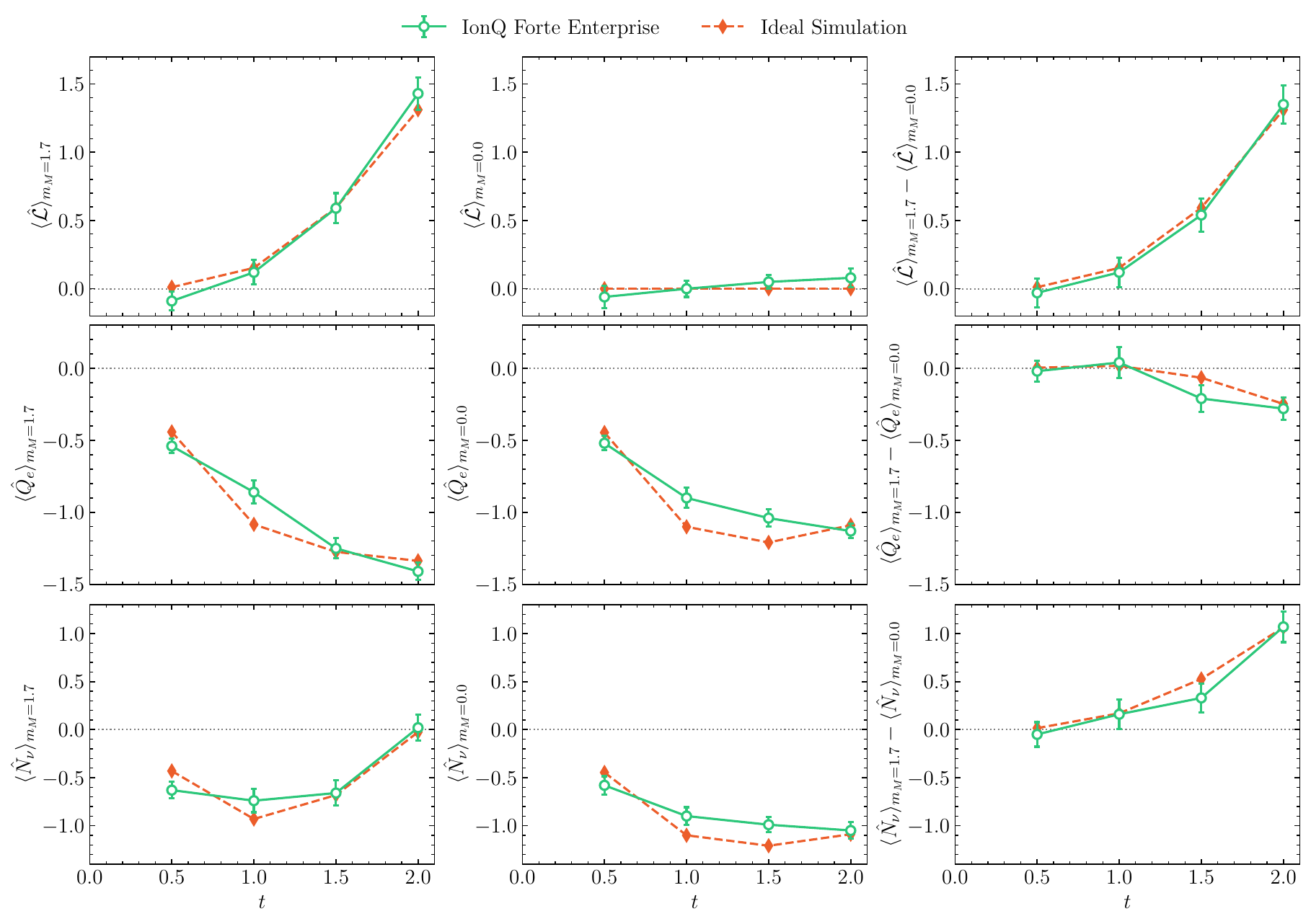}
\caption{The time evolution of the lepton number (upper row), 
lepton electric charge (middle row) and neutrino number (lower row) during the decay of 
the $|\Delta^-\Delta^-\rangle$ two-baryon state
in 1+1D QCD. 
Two steps of first-order Trotterized time evolution using 
the (approximate) valence-fermion weak interactions are implemented, requiring 470 two-qubit gates. The left panels show the results obtained with a Majorana mass of $m_M=1.7$, the center panels show results for $m_M=0$, and the right panels show the differences between the $m_M=1.7$ and  $m_M=0$ results. The green points were obtained from IonQ Forte Enterprise, and the orange diamonds correspond to noiseless simulation. The error bars are obtained from bootstrap resampling and represent one standard deviation, and the gray dotted line is added for reference.}
\label{fig:FE_results}
\end{figure*}

\noindent
Results obtained from quantum simulations using IonQ Forte Enterprise are shown in Fig.~\ref{fig:FE_results} for $t=\{0.5,1.0,1.5,2.0\}$ and 
two values of the Majorana mass $m_M=\{0.0,1.7\}$, as well as in Table~\ref{tab:postprocess_data}.
For times $t=\{0.5,1.0,1.5\}$ these results were derived from 14,400 shots, 
while 24,000 were taken for time $t=2.0$.
Approximately $10\%$ of the shots survived
after filtering on leakage detection and conservation of color and total electric charges.
The error bars on these results were computed by bootstrap resampling, see Supplementary Note 4 for details.

\begin{table*}[!ht]
\begin{ruledtabular}
\begin{tabular}{ccccccccc}
 & \multicolumn{4}{c}{Ideal Simulation} & \multicolumn{4}{c}{QPU Results} \\ \\[-1em]
 \cmidrule(lr){2-5} \cmidrule(lr){6-9}
 \\[-1em]
 $t$ & 
 $\langle\hat{\mathcal{L}}\rangle_{m_M=0}$ & $\langle\hat{\mathcal{L}}\rangle_{m_M=1.7}$& $\langle\hat{Q}_e\rangle_{m_M=0}$ & $\langle\hat{Q}_e\rangle_{m_M=1.7}$ & $\langle\hat{\mathcal{L}}\rangle_{m_M=0}$ & $\langle\hat{\mathcal{L}}\rangle_{m_M=1.7}$& $\langle\hat{Q}_e\rangle_{m_M=0}$ & $\langle\hat{Q}_e\rangle_{m_M=1.7}$
 \\ \\[-1em]
 \midrule \\[-1em]
 $0.5$ & $0.0$ & $0.01$ & $-0.45$ & $-0.44$ & $-0.06\pm 0.08$ & $-0.09\pm 0.07$ & $-0.52\pm 0.05$ & $-0.54\pm0.05$ \\ \\[-1em]
 $1.0$ & $0.0$ & $0.15$ & $-1.10$ & $-1.08$ & $\phantom{-}0.00 \pm 0.06$ & $\phantom{-}0.12\pm0.09$ & $-0.90\pm0.07$ & $-0.86\pm 0.08$ \\ \\[-1em]
 $1.5$ & $0.0$ & $0.59$ & $-1.20$ & $-1.26$ & $\phantom{-}0.05\pm0.05$ & $\phantom{-}0.59\pm0.11$ & $-1.04\pm0.06$ & $-1.25\pm 0.07$ \\ \\[-1em]
 $2.0$ & $0.0$ & $1.31$ & $-1.09$ & $-1.34$ & $\phantom{-}0.08\pm 0.07$ & $\phantom{-}1.43\pm 0.12$ & $-1.13\pm 0.05$ & $-1.41\pm 0.06$ \\ \\[-1em]
 \end{tabular}
\end{ruledtabular}
\caption{
The lepton number and electric charge obtained from a noiseless statevector simulator 
(Ideal Simulation) and from Forte Enterprise (QPU results). 
The QPU results correspond
to the green points (mean value and $\pm 1\sigma$ uncertainty) in Fig.~\ref{fig:FE_results}.
Non-linear filtering has been applied to the raw results, which are also post-selected on the conservation of total electric charge, red, green and blue color charges, as well as no leakage detected by the flag qubits.
}
\label{tab:postprocess_data}
\end{table*}

A clear (statistically significant) signal 
for the violation of
lepton number is 
found for $m_M=1.7$, as shown in Fig.~\ref{fig:FE_results},  which is absent for $m_M=0$.
Specifically, at $t=2.0$, there is a $10\sigma$ difference between the lepton numbers obtained with $m_M=1.7$ and $m_M=0$.
The ideal simulation results obtained from noiseless simulation using classical computers are also shown in Fig.~\ref{fig:FE_results}, and are in excellent agreement with the results obtained from Forte Enterprise. 
The effects of the different error mitigation methods applied during the simulations are shown in 
Supplementary Table 5 in Supplementary Note 5.
The impact of the combined error-mitigation on the raw results obtained from the device are essential in obtaining reliable results 
(as determined by comparison with the classically computed counterparts).
The flag gadgets do not make a statistically significant difference to the results, changing central values by less than $1\sigma$ at each time step.
On the other hand, post-selection is essential.

The lepton electric charge and lepton number in Fig.~\ref{fig:FE_results} (and extended to later times in Supplementary Note 2) do not exhibit the expected time dependence of an exponential decay. 
This is a finite-size effect, and an exponential decay is expected to emerge in the continuum and infinite volume limits, where the density of states becomes sufficiently dense at the kinematics of the transition energy.
This was studied in detail in Appendix D of Ref.~\cite{Farrell:2022vyh}.

For $m_M=0$, while the lepton number remains consistent with zero at all times, as expected, electric charge is produced in the lepton sector.  
This is consistent with 
single $\beta$-decay, $2\nu\beta\beta$-decay and transitions to other intermediate states that are energetically disfavored.
The baryons  and leptons cannot separate after a decay due to the small simulation volume, and the continual interactions between baryons  and leptons causes there to be a non-zero lepton electric charge density at all times.

The time dependence of lepton number and lepton electric charge, as displayed in Fig.~\ref{fig:FE_results}, 
could, in principle, 
reveal aspects of the underlying mechanism of the decay process.
Initially, the lepton sector has the quantum numbers of the vacuum, 
${\cal L}=Q_e=0$.
The expectation value of these quantities change in time due to the decay of the $|\Delta^- \Delta^-\rangle$ di-baryon.
For simplicity, consider
the two possible time-orderings of  two single $\beta$-decays and one Majorana mass term $\hat{H}_{\rm Maj}$,
the minimal operator structure required for $0\nu\beta\beta$-decay to occur. 
There will be contributions from an arbitrary number of insertions of these operators during unitary time evolution. 
However, in nature, contributions from 
multiple insertions are
 suppressed by the smallness of the weak coupling constant and Majorana mass.
In terms of the time-ordered sequence of interaction terms:
\begin{enumerate}
    \item 
    $\hat{H}_\beta^{1+1} $, then $\hat{H}_{\rm Maj}$, then $\hat{H}_\beta^{1+1} $:\\
    ${\hat Q}_e=0$, ${\cal L}=0$, then ${\hat Q}_e=-1$, ${\cal L}=0$, then ${\hat Q}_e=-1$, ${\cal L}=2$, 
    then ${\hat Q}_e=-2$, ${\cal L}=2$.
    \item 
    $\hat{H}_\beta^{1+1} $, then $\hat{H}_\beta^{1+1}$, then $\hat{H}_{\rm Maj}$:\\
    ${\hat Q}_e=0$, ${\cal L}=0$, then ${\hat Q}_e=-1$, ${\cal L}=0$, then ${\hat Q}_e=-2$, ${\cal L}=0$, 
    then ${\hat Q}_e=-2$, ${\cal L}=2$.   
\end{enumerate}
Note that the sequence with $\hat{H}_{\rm Maj}$ acting first does not contribute because the initial state is an eigenstate of the Hamiltonian without $\hat{H}_\beta^{1+1}$.
These, and higher order, reaction pathways interfere during the decay process
and need to be coherently summed together to predict
${\cal L}$ and ${\hat Q}_e$. 
This coherent evolution of all possible reaction pathways is one of the principle advantages of using quantum computers.
Given strong interaction times scales, it is clear that wavefunction evolution at the yocto-scale has the potential to
provide key information that can be used to identify dominant decay pathways.

\section{Discussion}
\label{sec:Summ}
\noindent
We have performed a suite of 
path-finding quantum simulations of $0\nu\beta\beta$-decay in 1+1D QCD
induced by a lepton-number violating Majorana neutrino mass. 
The simulations were performed on two lattice sites with two-flavors of quarks and leptons, which maps to 32 qubits, and with unphysical values of the Hamiltonian parameters.
Multiple facets of our simulations, from circuit design to error mitigation,
were co-designed to 
maximize the
performance from IonQ's Forte-generation trapped-ion quantum computers.
This included using extra qubits as ancillae to detect leakage on the most important qubits, and optimizing circuits for Forte’s all-to-all connectivity and native $R_{ZZ}$ two-qubit gates. 

A production run on IonQ Forte implementing the full weak operator requiring 2,356 two-qubit gates
established the limits of the initial co-design and helped to identify effective error-mitigation strategies. Informed by the results of these computations, circuits with 470 two-qubit gates were executed on Forte Enterprise achieving sufficient fidelity
to establish a $10\sigma$ signal for the dynamical generation of lepton number violation in $0\nu\beta\beta$-decay induced by the valence weak operator.
This demonstrated the neutrinoless decay mechanism of a two-baryon initial state that was only possible due to a non-zero neutrino Majorana mass.

Our work 
establishes a set of benchmarks for further 
quantum simulations of exotic weak decays
that may eventually impact experimental searches for new physics.
Simulating doubly-weak decay processes is a significant challenge for classical computing due to the necessity of coherently tracking the dynamics in a strongly-interacting nucleus.
Quantum simulations are expected to eventually improve upon results obtained with classical computation, 
and provide insight into the underlying strong-interaction mechanisms and pathway(s)
at yocto-second and longer time scales.

The next generation of quantum simulations will benefit from improvements in both fidelity and qubit count expected in the next few years~\cite{IonQ_roadmap2025}. 
Notably, by the end of 2027, quantum error correction  is anticipated to be implemented on IonQ's devices with $\approx 10000$ physical qubits and $\approx 800$ logical qubits. Today's quantum error correction on trapped ion qubits exhibits gate infidelities of as good as $\approx 5 \times 10^{-6}$ for Clifford gates \cite{Paetznick:2024ztu} and gate infidelities as good as $\approx 5 \times 10^{-4}$ \cite{daguerre2025experimental} for non-Clifford gates, but IonQ's roadmap aims for logical gate infidelities as good as $10^{-7}$ or less.
These advancements will enable progress toward more realistic simulations of exotic weak decays
that are required to begin making contact with experiment,
complementing an analogous path for lattice QCD combined with effective field theory that has been outlined in Refs.~\cite{Tiburzi:2017iux,Shanahan:2017bgi,Cirigliano:2020yhp,Nicholson:2018mwc,Tuo:2019bue,Davoudi:2020xdv,Davoudi:2020gxs,Detmold:2022jwu,Davoudi:2024ukx}.

Next steps along this path are to extend these simulations to 2+1D, building upon recent progress on efficient Hamiltonian formulations of lattice gauge theories~\cite{Grabowska:2024emw,Kadam:2024zkj,Illa:2025dou,Balaji:2025afl},
to increase the size of the spatial volumes using larger registers of qubits, to work closer to the continuum limit, 
to work with fermion masses closer to their experimental values,
and to perform simulations using a range of neutrino masses.
In the nearer-term, first steps toward the aforementioned achievements should be feasible with the release of the next-generation IonQ device, Tempo, and the repetition of this study with the full, unapproximated 1+1D QCD and decay Hamiltonian should be feasible shortly afterward.
Access to larger 1+1D lattices will allow the leptons emitted during 
$0\nu\beta\beta$-decay
to fully separate from the nucleus, which is important for mitigating finite size effects. 

As these quantum simulations become more sophisticated, 
it will be important to robustly quantify lattice spacing artifacts and
extrapolate to infinite volume and physical parameters.
Our results illuminate some of the challenges that lie ahead in approaching the physical hierarchy of mass scales that contribute to this type of decay process.   The physical strong-interaction energy scales are in the GeV region, while the neutrino masses are sub-eV. 
Naive methods of simulation will not be possible, even with quantum computers, simply because of the huge discrepancy in the relative time-scales that are involved.  
Similarly, another challenge will be the robust preparation of the initial-state nucleus in a large spatial volume. 
This is a difficult problem, even in classical lattice QCD simulations~\cite{NPLQCD:2012mex,Yamazaki:2015asa,Davoudi:2020ngi}, 
due to the excitation energies of nuclei being orders of magnitude smaller than their mass.

Despite these challenges, there are still several advantages expected from quantum simulations.
One is to provide insight into the underlying strong-interaction mechanisms and pathway(s) at yocto-second and longer time scales.
Once identified, these insights could motivate new, highly efficient, approximation schemes that would improve traditional calculations using classical computers.
Additionally, quantum simulations can be performed with
a selection of parameters away from the physical point, where the hierarchy between the neutrino mass and strong-interaction scale is not so severe.
With a sufficiently large ensemble of simulations, robust extrapolations to the physical parameters will be possible.

\section{Methods}

\subsection{The Simulation Setup}
\label{sec:Hami}
\noindent 
In our simulations, fermions are placed on the lattice using a staggered discretization that maps $L$ spatial sites to $N=2L$ staggered sites and $2LN_s$ fermion sites, where $N_s$ is the number of fermion species (for more details, see Supplementary Note 3).
The Hamiltonian that is used has four terms,
\begin{align}
\hat{H} =  \hat{H}_{\text{free}} + \hat{H}
_{\text{glue}}  + \hat{H}_{\beta, {\rm valence}}^{1+1} + \hat{H}_\text{Maj}\ .
\label{eq:H}
\end{align}
The Hamiltonian describing free staggered fermions is~\cite{Kogut:1974ag,Banks:1975gq}
\begin{align}
    \hat{H}_{\text{free}} = \sum_{f}\sum_{n=0}^{2L-1} & \left [m_f (-1)^n\phi^{(f)\dagger}_{n}\phi_{n}^{(f)} \right. \nonumber \\
    & \left. + \ \frac{1}{2}\left (\phi^{(f)\dagger}_{n} \phi_{n+1}^{(f)} + {\rm h.c.} \right )\right ]  \ ,
    \label{eq:Hfree}
\end{align}
where $m_f$ are the bare (Dirac) fermion masses with $f\in \{u_r,u_g,u_b,d_r,d_g,d_b,e,\nu\}$ corresponding to red, green and blue up and down quarks, and the two leptons, electrons and neutrinos. The index $n$ labels the staggered site, with $n$ even corresponding to fermion sites (quarks and leptons), and $n$ odd to anti-fermion sites (anti-quarks and anti-leptons).

The QCD interactions are encoded in $\hat{H}_{\text{glue}}$.
Explicit gauge degrees of freedom are 
absent in axial gauge~\cite{Farrell:2022wyt,Atas:2022dqm}, leaving a non-linear color Coulomb interaction between the quarks. 
With PBCs, the interaction is~\cite{Dempsey:2022nys}
\begin{equation}
\hat{H}_{\text{glue}}  = 
\frac{g^2}{2}
\sum_{n=0}^{2L-1}\sum_{s=1}^{\lambda}\left (-s + \frac{s^2}{2L} \right )(1-\frac{1}{2}\delta_{s,L}) \sum_{a=1}^8 
Q_n^{(a)} Q_{n+s}^{(a)} 
\  ,
\label{eq:HGaussLaw}
\end{equation}
where $g$ is the QCD interaction strength and 
\begin{equation}
Q_n^{(a)} \ = \ \sum_{f=u,d}\phi_n^{(f)\dagger}T^{(a)}\phi_n^{(f)}
\  ,
\label{eq:su3charges}
\end{equation}
are the eight $SU(3)$ charges (color indices have been suppressed). 
The dynamics of the zero-mode of the gauge field is not considered in this work, 
see Supplementary Note 3.
Recent work has shown that the mechanism of confinement motivates an approximate interaction of these naively long-range interactions~\cite{Farrell:2024fit}.
This approximation truncates interactions beyond $\lambda$ staggered sites, and converges exponentially for $\lambda$ larger than the confinement scale.
For the simulation parameters selected in this work, $\lambda=1$ is found to be sufficient, and will be used throughout.

Weak interactions are modeled through a local vector-like four-Fermi operator~\cite{Farrell:2022vyh},

\begin{equation}
   \hat{H}_{\beta, {\rm valence}}^{1+1} =
    \frac{G}{\sqrt{2}}  \sum_{n \ \text{even}}\left(
    \phi_n^{(u)\dagger}\phi_n^{(d)}\phi_n^{(e)\dagger}\phi_{n+1}^{(\nu)}
         + {\rm h.c.} \right) 
         \ ,
\label{eq:HbetaC1}
\end{equation}
where $G$ is the weak coupling constant (Fermi's constant).
This is an early-time approximation of the full four-Fermi operator that retains only the terms acting on the ``valence" quarks and leptons~\cite{Farrell:2022vyh}. 
Similar operator truncations were used in early quenched lattice QCD calculations, e.g.,
Ref.~\cite{Gockeler:2001xw}.
The impact of retaining terms beyond the valence sector are considered in 
Supplementary Note 3C.
Note that color indices,
which are summed over,
have been suppressed.
This approximation will be used to reduce the number of two-qubit gates required for time evolution (see Supplementary Note 2C).

Lastly, the neutrino Majorana mass term is~\cite{Farrell:2022vyh} 
\begin{align}
\hat{H}_{\rm Maj} =
\frac{1}{2} m_M 
     \sum_{n \ \text{even}} 
\left( 
\phi_n^{(\nu)}
\phi_{n+1}^{(\nu)}
+ {\rm h.c.}
\right) \ .
\end{align}
This is the unique local fermionic operator that violates lepton number by two units while preserving the other symmetries.

\begin{figure*}[!t]
    \centering
    \includegraphics[width=0.95\textwidth]{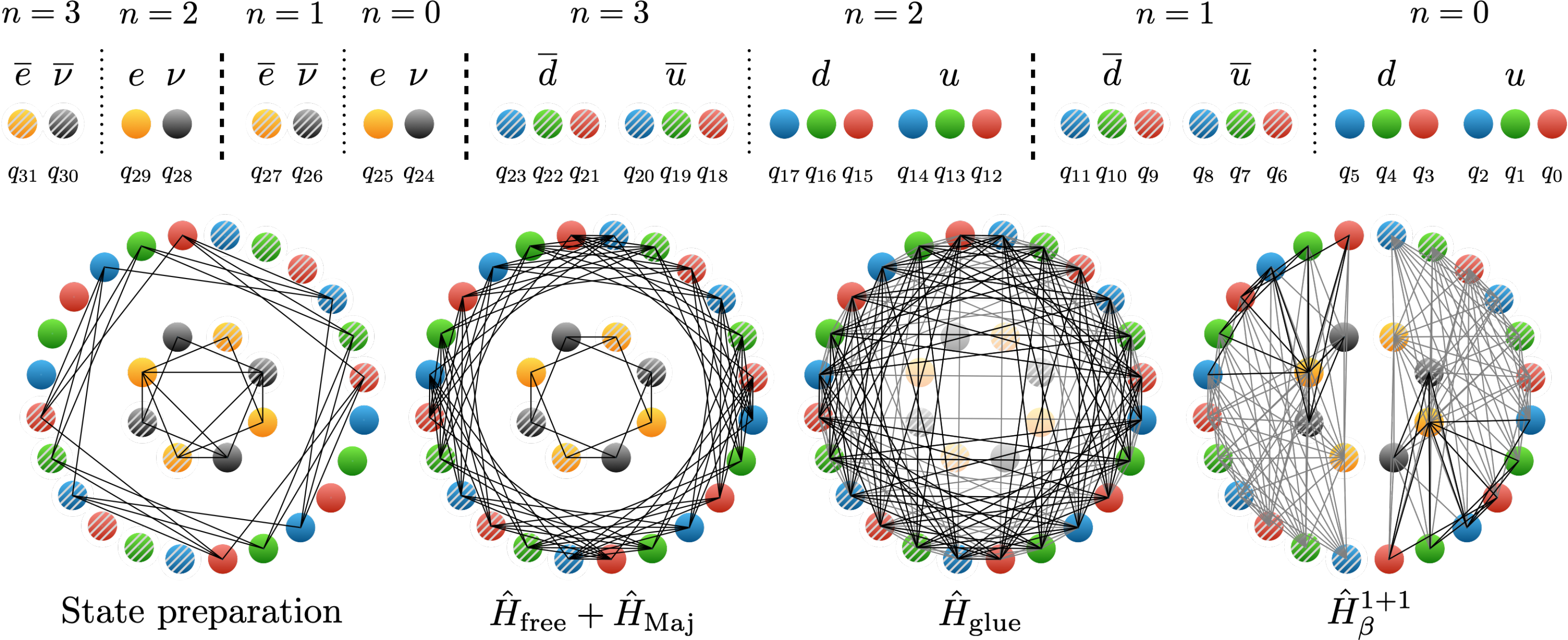}
    \caption{On top, the $L=2$ lattice-to-qubit mapping used in the $0\nu\beta\beta$-decay simulations on IonQ's quantum computers.
    Spatial sites are separated by dashed lines, while staggered fermion sites ($n=0,1,2,3$) are separated by dotted lines.
    Qubits are indexed from right-to-left, with the rightmost qubit being qubit 0, $q_0$.
    Qubits $q_0,q_1,...,q_{23}$ correspond to up ($u$) and down ($d$) quarks, each of three colors $r,g,b$.
    Qubits $q_{24},q_{25},...,q_{31}$ correspond to the electrons ($e$) and neutrinos ($\nu$).
    The ordering of the lepton and quark spatial sites has been chosen to reduce the length of Pauli strings in the Jordan-Wigner mapping.
    The bottom shows the required connectivity for state preparation and the Trotterized time evolution for each part of the Hamiltonian. 
    The gray connections are removed after the approximations described in the main text.}
    \label{fig:L2Qubits}
\end{figure*}

The Jordan-Wigner transformation (JW)~\cite{jordan:1928wi} is used to map the fermionic Hilbert space to $16L$ qubits.
The simulations in this work will be performed on $L=2$ spatial sites, with the qubit-to-lattice layout shown in Fig.~\ref{fig:L2Qubits}.
All of the terms in $\hat{H}$ are local on the level of spatial sites, but can become long-range operations when mapped to qubits. The non-trivial connectivity required to perform all the steps in the simulation is shown in the lower panels of Fig.~\ref{fig:L2Qubits}.
Because of this, the native all-to-all connectivity hosted by IonQ trapped-ion quantum computers is essential for efficient simulations.
The Hamiltonian expressed in terms of spin operators is given in Supplementary Note 3.

Hadronic weak decays begin with hadrons in the QCD sector and the vacuum in the lepton sector.
We choose to initialize a (maximally isospin stretched) two-baryon state, ``the $\Delta^-\Delta^-$ dibaryon",
\begin{align}
\vert \psi_{\text{init}} \rangle \ &= \
\vert \psi^{(\text{lep})}_{\text{vac}} \rangle \, \vert\Delta^-\Delta^-\rangle \ .
\label{eq:psiInit}
\end{align}
For $L=2$, the quark wavefunction factorizes into a tensor product of the one-flavor up-quark vacuum and a fully occupied $d$-quark register~\cite{Farrell:2022vyh}.
This factorization simplifies the quantum circuits needed for state preparation.

Current experimental searches for $0\nu\beta\beta$-decay involve atoms where $\beta$-decay is kinematically forbidden, but $\beta\beta$-decay is allowed.
This hierarchy occurs naturally in, for example, $^{76}\text{Ge}$, and produces a clear signal of $\beta\beta$-decay.
In our model, we reproduce this by tuning the masses and coupling constants to engineer the desired hierarchy in the spectrum.
Specifically, our simulations use
\begin{align}
& m_u = 1 \ , \ m_d  = 1.5 \ ,\nonumber \\
& m_e = 0.1 \ , \  m_{\nu} = 1.5 \ , \ m_M = \{0.0, 1.7\} \ , \nonumber \\ 
& g=1 \ , \ G=1 \ .
\label{eq:params_main}
\end{align}
This choice of parameters makes $\beta$-decay  energetically disfavored,
while the lepton number violating decay 
\begin{align}
\vert\Delta^-\Delta^-\rangle \ \to \ \vert\Delta^0\Delta^0\rangle + 2e^{-}
\end{align}
is not hindered by kinematical constraints.
This is the process we identify in our quantum simulations.
Additional details on the spectrum are given in Supplementary Note 6.
We emphasize that the parameters defining the 1+1D Hamiltonian implemented in this work are not related to those in nature.  
They are selected to enable a model simulation of $\beta\beta$-decay that is a first step toward future simulations in this genre of 
fundamental physics.
Supplementary Note 7 provides a short discussion of future extrapolations that will be required to be able to make predictions for physical observables.

\subsection{Overview of IonQ's Trapped-Ion Quantum Computers}
\label{subsec:IonQ}
\noindent
In this work we made use of two of IonQ's Forte-generation quantum processing units (QPUs)~\cite{Chen2024-co}, Forte and Forte Enterprise. These systems are very similar, with Forte Enterprise being the second device to be manufactured and made available with the Forte architecture. In these systems, $36$ qubits are realized as trapped $^{171}$Yb$^+$ ions, with quantum information encoded in two hyperfine levels of the ground state.
Ions are sourced via laser ablation and selective ionization before being loaded into a surface linear Paul trap in a compact integrated vacuum package.
Manipulation of the qubit states is achieved by illuminating individual ions with pulses of 355~nm light that drive two-photon Raman transitions, thereby enabling the implementation of arbitrary single-qubit rotations and two-qubit $R_{ZZ}$ entangling gates.
The median direct randomized benchmarking (DRB)~\cite{Proctor:2019PRL} fidelities of the entangling gates at the time of execution were 99.3\% on Forte and 99.5\% on Forte Enterprise, with gate durations around 950 $\mu$s.

IonQ Forte-generation devices integrate acousto-optic deflectors (AOD) that allow for independent steering of each laser beam to its respective ion, substantially reducing beam alignment errors across the ion chain~\cite{Kim:2008ApOpt,Pogorelov:2021PRXQ}.
This optical architecture, combined with a robust control system that automates calibration and optimizes gate execution, has enabled the realization of larger qubit registers with enhanced gate fidelities.

\subsection{Circuit-Optimization and Error-Mitigation}
\label{subsec:EM}
\noindent
When working with a quantum computer with limited resources, it is important to design the 
quantum circuits and error handling in tandem.
On NISQ devices, this amounts to
careful circuit optimization and error mitigation techniques~\cite{cai2023quantum}.
Quantum error-mitigation techniques
make use of methods like symmetrization, twirling~\cite{PhysRevA.72.052326},
amplification and extrapolation, regression, and/or post-selection in order to arrive at an approximation of the ideal circuit outputs for a given application. 
Noise-induced biases can be removed by leveraging application and device symmetries, 
with a smaller overhead cost than with random sampling. 
This approach can be especially efficient when the device-level biases are known~\cite{symm2023}. 
Combining the careful selection of circuit-variant implementations with observable-specific post-selection rules allows for precise error detection and higher shot efficiency. 
The approach chosen here combines debiasing through symmetrization~\cite{symm2023}, post-selection on symmetry checks, and a novel parametrized nonlinear filtering method on the lepton qubit register. 
Post-selection is based on the usage of spare qubits for flag-based~\cite{Ken2020} mid-circuit symmetry checks and leakage error detection to further reduce the errors.

The all-to-all connectivity 
and native $R_{ZZ}(\theta)$ gates available on IonQ’s quantum computers are used
to further optimize quantum circuits by merging blocks of two-qubit gates.
This reduces the number of entangling-gates by 15\%. 
All-to-all connectivity also allows for an efficient reduction in the bias between the different circuit implementations by varying the qubit-to-ion assignment without any additional gates. In addition to qubit remapping, we make use of a type of phase-flip twirling of two qubit gates in generating our variants as described in Ref.~\cite{symm2023}. For this project we also introduced a bit-flip symmetrization of readout into this process. This symmetrization involves generating pairs of circuit variants, where one of the pair applies bit-flips before and after measurement on a random set of qubits. The other variant of the pair is identical except for having the bit-flips before and after measurement applied to the complement of the set of qubits chosen for the first variant.

In post-processing,
we combine the post-selected measurement statistics from different twirled variants, 
filtering out outlier bit strings.
This filtering is accomplished by checking if a given measured bit string appears in at least some specified number of variants, referred to as the filter threshold.
The choice of the threshold is determined by a combination of knowledge of the device noise, the number of twirled variants, and the number of shots taken per variant.
A higher threshold better mitigates hardware-noise induced biases, but requires more variants and shots per variant.
Indeed, for a fixed number of variants and shots, a threshold that is set too high will have 
simple finite sampling effects that can lead to the loss of all information. 

To further optimize circuit performance, 
we used multiple noise tailoring, error mitigation and error detection methods. 
Details of each method are
given in Supplementary Notes 5 and 8.
We apply Pauli twirling~\cite{Wallman:2015uzh}, 
XY4 dynamic decoupling~\cite{Viola:1998jx,Ezzell:2022uat} and measurement twirling through bit flipping~\cite{PhysRevA.105.032620,Smith_2021}.
These techniques mitigate coherent two-qubit over- and under-rotations, phase and idling errors, spontaneous emission and measurement biases.
Each circuit is compiled to pairs of twirled variants with unique qubit assignments to debias qubit-associated error dependencies. 
Each pair of variants is identical up to bit flips before and after measurement that symmetrize readout errors. 
For the circuits run on Forte Enterprise, a total of 96 twirled variants, each with 150 shots, was chosen to balance
the ability to mitigate (systematic) hardware errors by increasing the number of variants, and reducing statistical uncertainties by performing more shots per variant. For $t=2$, we included 64 additional variants, totaling 160 (24,000 shots total).
The tradeoff between number of twirled variants and shots per variant was informed by a Monte Carlo analysis over a uniform distribution of possible output bit string distributions, assuming the hardware error rates taken from benchmarking data. 
See Supplementary Note 9 for details.

Our simulations only utilized 32 of the 36 qubits available on the Forte-generation machines.
On Forte Enterprise, the remaining four qubits are used to flag qubit leakage outside of the computational subspace. 
See Supplementary Note 5 for more information.
Measured bit strings are post selected based on the ancillae states indicating the absence of leakage.
Additionally, bit strings are post-selected to conserve color and total electric charges.
The charge operators are given in Supplementary Note 3.
On Forte, the 4 extra qubits on the 36 qubit register were used for mid-circuit symmetry iSWAP checks.
See Supplementary Note 5 for more details.

\subsection{Error mitigation strategies informed from Forte runs}
\label{sec:Forte_codesign}
\noindent
To determine the non-linear filtering threshold, number of variants and 
the implementation of mid-circuit symmetry checks, smaller test circuits with four-qubit Givens rotations were run.
The non-linear filtering threshold was adjusted to have no fewer than 200 surviving counts after filtering.
See Supplementary Note 8 for more details.
After analyzing the results from the runs on Forte, we noticed that leakage errors make a large contribution. 
This informed our decision to perform leakage checks instead of iSWAP checks on our second set of runs on Forte Enterprise.
For the second set of runs,
we kept the same filtering threshold and added post-selection on conserved charges:
$r=b=g=2$ and the total electric charge $Q_{\text{tot}}=-2$.

On Forte, we ran
reference circuits similar to the ones used in the operator decoherence renormalization (ODR) error-mitigation strategy~\cite{Farrell:2023fgd,Farrell:2024fit,Urbanek_2021,ARahman:2022tkr,Farrell:2022wyt,Ciavarella:2023mfc}.
The output of these reference circuits can be efficiently determined using classical computing, and deviations from expected results are used to learn features of the device noise.
However, due to the limited number of shots and residual bias in the single-qubit $\hat{Z}$ observables, we could not apply ODR to the 2,356 two-qubit gate circuits. 
Instead, we used the reference circuits to inform our noise models.
For these experiments, instead of ODR we used debiasing with non-linear filtering, which we found to be more effective
with noisier results and smaller numbers of shots. 
Further, the non-linear filtering works well with post-selection, 
while ODR is not compatible with it.
The information gathered from these limit-testing runs will be valuable in designing the next generation of $0\nu\beta\beta$-decay simulations on trapped-ion platforms.

\subsection{Quantum circuits for simulating weak decays}
\label{sec:classSim}
\noindent
The initial state of our simulations is $|\psi_{\text{init}}\rangle=\vert \psi^{(\text{lep})}_{\text{vac}} \rangle \, \vert\Delta^-\Delta^-\rangle $.
Preparation of the lepton vacuum is straightforward, as it is two flavors of non-interacting fermions~\cite{Wecker_2015,Jiang:2017pyp,Kivlichan_2018}.
As mentioned above, the quark wavefunction factorizes between the $u$ and $d$ sectors as
\begin{align}
\vert\Delta^-\Delta^-\rangle \ = \ |\psi^{(u)}_{\text{vac}}\rangle|0\rangle^{\otimes 6} \ .
\end{align}
The $|0\rangle^{\otimes 6}$ represents the fully occupied register of $d$ quarks and is trivial to prepare.

\begin{figure*}[!th]
    \centering
    \includegraphics[width=0.95\textwidth]{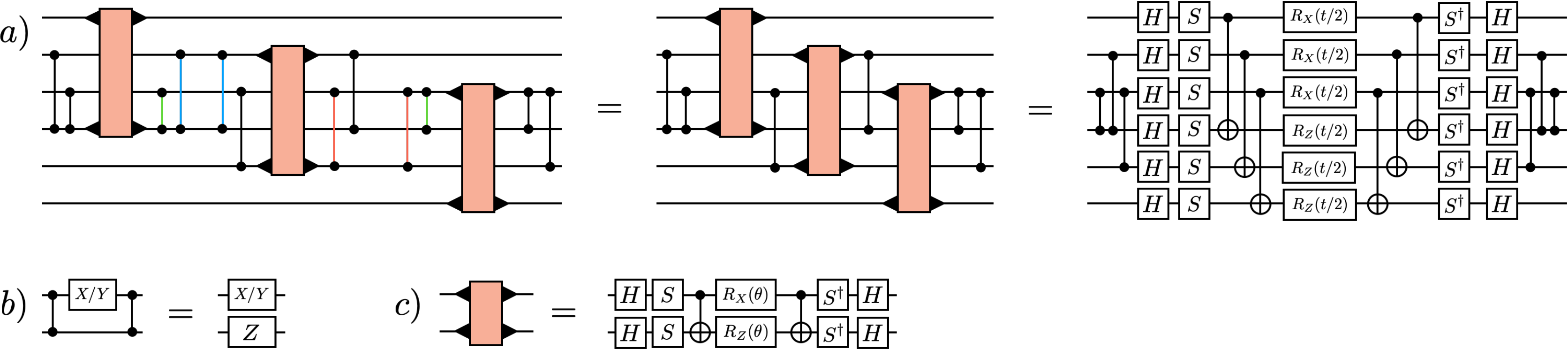}
    \caption{A method for constructing shallow circuits on quantum computers with all-to-all connectivity.
    The barbells denote  $CZ$ gates, and the 
    \rotatebox[origin=c]{-90}{$\blacktriangle$}-symbols 
    on the orange blocks mark the qubits that are acted on.
    a) A quantum circuit that implements the time-evolution of the one-flavor quark kinetic term in Eq.~\eqref{eq:HkinNf1_main}. 
    The red, blue and green $CZ$ gates cancel against each other.
    b) A useful circuit identity.
    c) The definition of the light orange circuit block that implements 
    $e^{-i\theta( \hat{\sigma}^+\hat{\sigma}^-+\hat{\sigma}^-\hat{\sigma}^+)}$.}
    \label{fig:CZExample}
\end{figure*}

To prepare $|\psi^{(u)}_{\text{vac}}\rangle$, we use the Scalable Circuit ADAPT-VQE~\cite{Grimsley_2019} (SC-ADAPT-VQE) algorithm developed in Refs.~\cite{Farrell:2023fgd,Farrell:2024fit}.
SC-ADAPT-VQE is a variational quantum algorithm that utilizes symmetries and hierarchies in length scales to determine shallow state preparation circuits.
On a two spatial site lattice, the SC-ADAPT-VQE ansatz consists of
a single parameterized circuit $e^{i \theta \hat{O}}$ that is real and preserves the symmetries of the QCD Hamiltonian.
An operator with these properties is constructed from the commutator of the mass and kinetic terms (for the $u$ quarks) in $\hat{H}_{\text{free}}$,
\begin{align}
\hat{O} &= i \sum_{n=0}^{2L-1}\left [(-1)^n\phi_n^{(u)\dagger}\phi_n^{(u)} , \left (\phi^{(u)\dagger}_{n} \phi_{n+1}^{(u)} + {\rm h.c.} \right )
\right ] \ .
\end{align}
The variational optimization and the construction of the associated quantum circuits is discussed at length in Supplementary Note 10.

Next, the decay process is simulated by time evolving the initial state with $e^{-i \hat{H} t}$ using a first-order Trotterization.
By judiciously ordering the terms in the Trotter decomposition, 
the first Trotter step can be simplified using 
$(\hat{H}_{\text{free}} + \hat{H}_{\text{Maj}}+\hat{H}_{\text{glue}})|\psi_{\text{init}}\rangle = E_{\text{init}} |\psi_{\text{init}}\rangle$~\cite{Farrell:2022vyh}.
We introduce a new method for constructing the required time-evolution quantum circuits that builds on results presented in Ref.~\cite{Chernyshev:2025jyw}.
These circuits feature a high degree of parallelizability when compiled to a device with all-to-all connectivity.
The circuit construction will briefly be described here, with more information in Supplementary Note 10.
An obstacle in the way of highly parallelizable circuits is the JW transformation, which adds a string of Pauli-$\hat{Z}$ operators to all non-mass terms in the Hamiltonian.
This can be overcome by first designing the circuits without the JW $\hat{Z}$ strings, and then including them by conjugating the $\hat{Z}$-string-free circuits by a sequence of $CZ$s, taking advantage of the identity shown in Fig.~\ref{fig:CZExample}b).

For demonstration, consider the time-evolution of a kinetic term for one flavor of quark over one spatial site,
\begin{align}
\hat{H}_{\text{kin}} \ = \ \frac{1}{2}\sum_{c=0}^{2}\left [\hat{\sigma}^+_c \hat{Z}_{c+1} \hat{Z}_{c+2} \hat{\sigma}^-_{c+3} \: + \: {\rm h.c.} \right ] \ .
\label{eq:HkinNf1_main}
\end{align}
The steps for creating a circuit that implements the unitary evolution of $e^{-it\hat{H}_{\text{kin}}}$ are shown in Fig.~\ref{fig:CZExample}a). 
In the first equality, the fact that $(CZ)^2 = \hat{I}$ (highlighted with the same color) and that $CZ$s commute with each other has been used.
The second equality decomposes the orange box and pushes the $CZ$s to the beginning and end of the circuit.
This strategy is used to construct all of the time-evolution circuits in this work.
This circuit construction method generalizes ideas used in fermionic SWAP networks \cite{PhysRevA.71.032310,CerveraLierta2018exactisingmodel, Kivlichan_2018}. This is because the fermionic SWAP gate is equivalent to a qubit SWAP operation followed by a $CZ$ gate, the former of which can be implemented virtually with all-to-all connectivity.

\section*{Data Availability}
\noindent
The data that support the findings of this study are available from the
corresponding author upon request.

\bibliography{bib_main}

\vskip 0.1in

\begin{acknowledgments}
\noindent
We would like to thank Saurabh Kadam, Joe Latone, and Torin Stetina for helpful discussions and support.
This work was supported, in part, 
by U.S.\ Department of Energy, Office of Science, Office of Nuclear Physics, InQubator for Quantum Simulation (IQuS) under Award Number DOE (NP) Award DE-SC0020970 via the program on Quantum Horizons: QIS Research and Innovation for Nuclear Science
(MJS, IC, RCF),
and 
by the Quantum Science Center (QSC) which is a National Quantum Information Science Research Center of the U.S.\ Department of Energy (MI, IC).
This work is also supported in part by Los Alamos National Laboratory, which is operated by Triad National Security, LLC, for the National Nuclear Security Administration of U.S. Department of Energy (Contract No. 89233218CNA000001) (IC).
This work is also supported, in part, through the Department of Physics and the College of Arts and Sciences at the University of Washington.
RCF acknowledges support from the U.S. Department of Energy QuantISED program through the theory consortium “Intersections of QIS and Theoretical Particle Physics” at Fermilab, from the U.S. Department of Energy, Office of Science, Accelerated Research in Quantum Computing, Quantum Utility through Advanced Computational Quantum Algorithms (QUACQ), and from the Institute for Quantum Information and Matter, an NSF Physics Frontiers Center
(PHY-2317110). 
RCF additionally acknowledges support from a Burke Institute prize fellowship.
This research used resources of the National Energy Research Scientific Computing Center (NERSC), a Department of Energy Office of Science User Facility using NERSC award NP-ERCAP0032083.
\end{acknowledgments}

\vspace{1cm}
\section*{Author contributions}
\noindent

{\bf Andrew Arrasmith}
designed and implemented the tool used in allocating the shot count per variant and number of variants;
co-developed and co-implemented the bootstrapping tools used to estimate the error bars reported;
analyzed and post-processed results from IonQ Forte and Enterprise; 
participated in bi-weekly meetings;
co-wrote the manuscript.

{\bf Aharon Brodutch}
participated in bi-weekly meetings; co-wrote the manuscript. 

{\bf Ivan Chernyshev} 
co-conceived the project and model Hamiltonian; 
co-developed the quantum circuits; 
performed classical simulations using exact diagonalization and statevector simulators; 
analyzed post-processed results from IonQ Forte and Enterprise; 
participated in bi-weekly meetings;
co-wrote the manuscript.

{\bf Roland Farrell} 
co-conceived the project and model Hamiltonian; 
co-developed the quantum circuits; 
performed classical simulations using exact diagonalization and statevector simulators; 
analyzed post-processed results from IonQ Forte and Enterprise; 
participated in bi-weekly meetings;
co-wrote the manuscript.

{\bf Claudio Girotto}
participated in bi-weekly meetings; co-wrote the manuscript. 

{\bf Marc Illa} 
co-conceived the project and model Hamiltonian; 
co-developed the quantum circuits; 
performed classical simulations using exact diagonalization and statevector simulators; 
analyzed post-processed results from IonQ Forte and Enterprise; 
participated in bi-weekly meetings;
co-wrote the manuscript.

{\bf Ananth Kaushik} led the technical execution of the project and co-led the coordination of the activities of the IonQ and UW teams; executed the quantum circuits on the IonQ QPUs IonQ Forte and Enterprise machines; analyzed and post-processed results from hardware; 
participated in bi-weekly meetings;
co-wrote the manuscript.

{\bf Miguel Angel Lopez-Ruiz}
analyzed and post-processed the results from hardware runs; participated in bi-weekly meetings; co-wrote the manuscript. 

{\bf Andrii Maksymov} co-designed and co-conceived the error mitigation and error detection techniques for IonQ Forte and Enterprise machines; co-developed the quantum circuits; analyzed and post-processed results; participated in bi-weekly meetings; co-wrote the manuscript. 

{\bf Martin Roetteler}
co-led and co-conceived the project;  
analyzed post-processed results from IonQ Forte and Enterprise; 
participated in bi-weekly meetings;
co-wrote the manuscript.

{\bf Martin Savage} 
co-led and co-conceived the project; co-developed the model Hamiltonian; 
verified the quantum circuits; 
analyzed post-processed results from IonQ Forte and Enterprise; 
participated in bi-weekly meetings;
co-wrote the manuscript.

{\bf Yvette de Sereville} co-developed software infrastructure for error mitigation, compilation and circuit execution on IonQ Forte and Enterprise; analyzed and post-processed results from IonQ Forte and Enterprise; participated in bi-weekly meetings; co-wrote the manuscript. 

{\bf Felix Tripier} co-designed and co-developed the error mitigation and error detection techniques for IonQ Forte and Enterprise machines; co-developed software infrastructure for error mitigation, compilation and circuit execution on IonQ Forte and Enterprise; co-developed the quantum circuits; analyzed and post-processed results; participated in bi-weekly meetings; co-wrote the manuscript.

\section*{Competing interests}
\noindent
The authors declare the following competing interests: 
AM, FT, MALR, AA, YDS, AB, CG, AK and MR are employees and equity holders of IonQ, Inc. MJS serves on an IonQ advisory committee and owns stock.
IAC, RCF and MI have no competing interests.

\clearpage
\onecolumngrid

\begingroup
\setcounter{section}{0}
\setcounter{figure}{0}
\setcounter{table}{0}
\renewcommand{\thesection}{Supplementary Note~\arabic{section}}
\renewcommand{\figurename}{Supplementary Figure}
\renewcommand{\thetable}{\arabic{table}}
\renewcommand{\tablename}{Supplementary Table}

\newcommand{\stab}{Supplementary Table}
\newcommand{\sfig}{Supplementary Fig.}
\newcommand{\seq}{Supplementary Eq.}

\begin{center}
    {\Large \bfseries Supplementary Information}
\end{center}
\vspace{1em}

\section{Some Relevant Background about \texorpdfstring{$0\nu\beta\beta$}{}-Decay}
\label{app:physback}
\noindent
In the Standard Model, fundamental symmetries prevent neutrinos from having a mass.
There is a tremendous amount of evidence that neutrinos are massive~\cite{Esteban:2024eli}, and an explanation requires physics beyond the minimal Standard Model.
One solution is that neutrinos are their own anti-particles~\cite{majorana1937,racah1937sulla}, 
allowing them to receive a Majorana mass contribution at the expense of violating lepton number.\footnote{
On general grounds, any lepton number-violating interactions will give rise to a Majorana mass through quantum fluctuations.}
A Majorana mass emerges naturally by considering the Standard Model as an effective description of nature that is valid below some high-energy scale.
One observable consequence of such a mass term is the $0\nu\beta\beta$-decay of 
certain nuclei~\cite{PismaZhETF.34.148}.
While $\beta\beta$-decays including the emission of two neutrinos 
($2\nu\beta\beta$)
proceed via a well-known (doubly-weak) mechanism that conserves lepton number,
$0\nu\beta\beta$-decays are forbidden in the  minimal Standard Model by lepton-number conservation.
This has motivated many ongoing and planned experimental programs that search for signatures of $0\nu\beta\beta$ decay~\cite{EXO-200:2019rkq,GERDA:2020xhi,CUORE:2021mvw,KamLAND-Zen:2022tow,
CUPID:2022puj,Majorana:2022udl,KamLAND-Zen:2024eml,AMoRE:2024loj,LEGEND:2025jwu,Adams:2022jwx,Agostini:2022zub,Barabash,Lewitowicz:2025qlr}.
One should keep in mind that a Majorana neutrino mass is not the only mechanism that can give rise to 
$0\nu\beta\beta$-decay.  
Contact interactions of higher dimension will generally be present if lepton number is violated, also contributing to the decay rate, 
e.g., Refs.~\cite{Mohapatra_1999,Savage:1998yh,Cirigliano:2022oqy}.
In either situation, low-energy physics may provide
a key window into physics at very high energies.

The $0\nu\beta\beta$-decay rates of nuclei are notoriously difficult to compute,
limiting interpretations of lifetime lower bounds predicted by theory and the implications of future potential experimental observations (for an early review, see Ref.~\cite{Haxton:1984ggj}).
The reasons for this difficulty range from the
many-body contributions to matrix elements, for example, 
as encapsulated in nuclear effective field theory 
(EFT)
treatments (e.g., see Refs.~\cite{Savage:1998yh,Prezeau:2003xn,Shanahan:2017bgi,physrevd.96.054505,PhysRevLett.120.202001,Cirigliano:2019vdj,Castillo:2024jfj}),
to the coherent evolution of excited states of the nucleus.
There have been recent theoretical advances in better understanding the Majorana neutrino decay mechanism in the context of
Euclidean-space lattice QCD (e.g., Refs.~\cite{Tiburzi:2017iux,Shanahan:2017bgi,Cirigliano:2020yhp,Nicholson:2018mwc,Tuo:2019bue,Davoudi:2020xdv,Davoudi:2020gxs,Detmold:2022jwu,Davoudi:2024ukx}), nuclear shell model (e.g., Refs.~\cite{Horoi:2015tkc,Iwata:2016cxn,Coraggio:2020iht,Jokiniemi:2021qqv,Jokiniemi:2022ayc}),
and EFT.
Specifically, the identification of KSW-type~\cite{Kaplan:1996xu,Kaplan:1998we,Kaplan:1998tg} logarithms from neutrino exchange~\cite{PhysRevLett.120.202001,Cirigliano:2019vdj}, leading to the promotion of multi-nucleon contact interactions in the 
EFT
power-counting from renormalization-group scaling.

To better understand aspects of these challenges, it is helpful to dissect the Majorana-mass induced decay mechanism. In the parent nucleus, one time-ordered pathway is that one neutron undergoes single-$\beta$ decay emitting an $e^-$ and a $\overline{\nu}_e$, transforming into a proton. The $\overline{\nu}_e$ converts into a $\nu_e$ via the Majorana mass term, which induces an inverse $\beta$-decay of another neutron. 
The (effective)
two-body $nn\rightarrow pp e^-e^-$ subprocess occurs within a nucleus, and 
the amplitudes of all such time-ordered pathways
must be summed over. 
Part of the complexities of this process come from the fact that 
nucleons in the nucleus are spatially overlapping, strongly interacting and in a correlated two-species fermion wavefunction.  
At a fundamental level, the $\beta$-decay processes are defined in terms of interactions with the quarks via the charged-current weak interaction.  
These quark-level interactions are evaluated in the nuclear wavefunction and are highly impacted by the correlations between nucleons. 
In the time-ordered pathway described above, the nucleus time-evolves from the 
ground state of the strong-interaction Hamiltonian for $N$ neutrons and $Z$ protons, to an intermediate state of the $(N-1,Z+1)$ Hamiltonian, to the ground state of the $(N-2,Z+2)$.
In the intermediate time region there are interferences between multiple nuclear energy levels that are
coupled with the electroweak Hamiltonian. 
In the interval of time between the first and second $\beta$-decay, the neutrino can be assumed to be freely propagating.
However, it is the complete dynamics of the coupled system that is challenging, and whose simulation is potentially better suited for quantum computers.

\section{Simulation Errors from Approximations}
\label{app:classical_simulations}
\noindent 
To reduce the depth of quantum circuits, the following approximations were implemented:
\begin{itemize}
    \item The initial $|\Delta^- \Delta^-\rangle$ 
    states were 
    prepared approximately using SC-ADAPT-VQE.
    Details are given in \ref{app:tevolCircs}, 
    and \stab~\ref{tab:AdaptEF} shows that the fidelity of the prepared state is $0.99992$.
    \item The range of the chromoelectric interactions is truncated to $\lambda=1$ staggered site.
    This is discussed in \ref{app:HamDetails}.
    \item Time evolution is implemented with $n_T=2$ first-order Trotter steps.
    \item The first Trotter step is simplified using the fact that $|\psi_{\text{vac}}^{(\text{lep})}\rangle|\Delta^-\Delta^-\rangle$ is an eigenstate of the Hamiltonian without $\hat{H}_{\beta}^{1+1}$.
    This will be discussed below.
    \item Pauli terms in the Hamiltonian with coefficients smaller than $t/(16 n_T)$ are removed from the time evolution unitary.
    This effectively removes single-qubit gates with small rotation angles.
    For the parameters chosen, this simplification only affects terms in the $\hat{H}_{\rm glue}$ of \seq~\eqref{eq:BetaHamL2_glue}.
    \item An approximate $\hat{H}_{\beta}^{1+1}$ interaction that only acts on the valence quarks is used.
    In our case, the $\beta$-decay process transforms $d\rightarrow u e \overline{\nu}$, therefore the terms acting on these fermions are labeled as valence terms, the rest are labeled as sea terms. Analogous to
    Ref.~\cite{Farrell:2022vyh}, 
    we approximate the interaction by only keeping the term that acts on the valence space, shown in \seq~\eqref{eq:HbetaC1}.
    This corresponds to the first term (and its hermitian conjugate) in \seq~\eqref{eq:BetaHamL2_beta}.
\end{itemize}
This supplementary note will compare the exact results obtained from exact diagonalization to results with different levels of approximation.

\subsection{Exact Diagonalization}
\label{app:exact}
\noindent
Simulations of $e^{-i\hat{H} t}|\psi_{\text{init}}\rangle$ using exact diagonalization provide
the exact result used to benchmark the impact of the various approximations.
Exact diagonalization can be performed
on fairly large systems by offloading some of the computation to the tensor product basis, and only constructing explicit matrices when necessary.
Enforcing symmetries at the level of the allowed basis states results in a much smaller effective Hilbert space.

As an example, consider a single flavor of quark in 1+1D QCD.
In the computational basis, each state is expressed as a string of $3N$ $0$s and $1$s, giving $2^{3N}$ states in total. 
Baryon number ${\cal B}$ separately constrains the occupation of the red ($r$), green ($g$) and blue ($b$) fermion sites to satisfy $r=g=b={\cal B}$.
This reduces the size of the Hilbert space to $\binom{N}{N/2-{\cal B}}^3$ states.
In addition, there are also spacetime symmetries and the remaining global $SU(3)$ (color singlet) constraints.
Consider the vacuum sector, which has momentum ${\bf k}=0$ and is even under charge conjugation ($C=+1$) and parity ($P=+1$).
These symmetries imply that certain bit strings will always 
contribute to
the wavefunctions with equal amplitudes, and with relative signs that can be determined from how the symmetries are realized.
The JW mapping makes the realization of symmetries subtle and is discussed in Ref.~\cite{Farrell:2025nkx}.
In addition, all physical states $|\psi\rangle$
are color singlets, i.e., they satisfy  $\sum_{n} \hat{Q}_n^{(a)} \vert \psi \rangle = 0$. 
Out of the eight charges $\hat{Q}^{(a)}$, the two diagonal constraints have already been enforced by selecting $r=g=b={\cal B}$, leaving six additional constraints.
One implication of being a color singlet is that, for example, global $SU(3)$ rotations of the form 
$\exp{i \frac{\pi}{2}\sum_n \hat{Q}^{(1)}_n}\vert \psi \rangle = \vert \psi \rangle$ leave the state invariant.
This transformation exchanges red and green quarks $r\leftrightarrow g$, while introducing a factor of ``$i$''.
These constraints, and the corresponding ones for 
$r\leftrightarrow b$ and $g\leftrightarrow b$, 
allow for larger sets of bit strings related by symmetry to be grouped together.\footnote{These constraints do not commute, and in practice only enforcing, e.g., $r\leftrightarrow g$ and $g\leftrightarrow b$ symmetry
leads to the smallest basis.}
The remaining non-Abelian constraints are difficult to enforce in the tensor product basis, and the complete color singlet basis can be found from determining the null space of the operator, $\hat{H}_{{\bf 1}} = ( \sum_nQ_n^{(a)} )^2$.
An efficient way to determine the null space is with the QR factorization of $\hat{H}_{{\bf 1}}$.

By building the Hamiltonian matrix in a basis that satisfies these constraints, operations such as diagonalization and matrix exponentiation become less computationally demanding.
To illustrate its utility, the size of the Hilbert space at different stages in the symmetry projection for the $L=5$ vacuum sector is
\begin{equation}
2^{30} \ \ \xLongrightarrow[]{r=g=b=0} \  \ 2^{24} \ \ \xLongrightarrow[r \leftrightarrow g, g\leftrightarrow b]{{\bf k}=0, C=P=+1} \ \ 2^{18} \ \ \xLongrightarrow[]{\hat{H}_{{\bf 1}}\vert \psi \rangle = 0} \ \  2^{15} \ ,
\end{equation}
rounding to the nearest power of $2$.

\subsection{Trotterized Time Evolution}
\label{app:trotter}

\noindent
The circuits presented in \ref{app:tevolCircs} act on 32 qubits, and can be simulated exactly with a state-vector simulator on a small cluster. The NVIDIA package CUDA-Q~\cite{The_CUDA-Q_development_team_CUDA-Q}, running on four A100 GPUs, was used to perform these simulations, together with Qiskit~\cite{Javadi-Abhari:2024kbf} to define the circuits.
As noted in our previous work~\cite{Farrell:2022vyh}, the terms in each Trotter step can be ordered so that the first step has the QCD Hamiltonian acting on an eigenstate.
This furnishes an overall phase, and therefore the first Trotter step need only have the term with $\hat{H}_\beta^{1+1}$.
Explicitly, for $n_T$ Trotter steps
in a leading-order (first-order) Trotter expansion,\footnote{More explicitly, the kinetic terms are broken up into even-odd staggered site and odd-even staggered site hoppings, and the free Hamiltonians are ordered mass then kinetic.}
\begin{align}
e^{-i\hat{H} t} \vert \psi_{\text{init}}\rangle \ &\approx  \left ( e^{-i t/n_T \hat{H}_{\beta}^{1+1}} e^{-i t/n_T \hat{H}_{\text{glue}}}e^{-i t/n_T \hat{H}_{\text{Maj}}}e^{-i t/n_T \hat{H}_{\text{free}}} \right )^{n_T} \vert \psi_{\text{init}}\rangle \nonumber \\[4pt]
&\approx 
\left ( e^{-i t/n_T \hat{H}_{\beta}^{1+1}} e^{-i t/n_T \hat{H}_{\text{glue}}}e^{-i t/n_T \hat{H}_{\text{Maj}}}e^{-i t/n_T \hat{H}_{\text{free}}} \right )^{n_T-1} e^{-i t/n_T \hat{H}_{\beta}^{1+1}} \vert \psi_{\text{init}}\rangle \ ,
\end{align}
with $\vert \psi_{\text{init}}\rangle = |\psi_{\text{vac}}^{(\text{lep})}\rangle|\Delta^- \Delta^-\rangle$.
In this work, the number of Trotter steps $n_T$ is fixed for all the times $t$.
The quantum simulations run on the IonQ Forte-generation QPUs used $n_T=2$. 
To quantify the Trotter errors, a comparison between the results from exact diagonalization and $n_T\in\{1,\ldots,64\}$ is shown in \sfig~\ref{fig:Trotter_results}, for the lepton number, $\hat{\mathcal{L}}$, and the electric charge in the lepton sector, $\hat{Q}_e$, defined in Eq.~1 in the main text.
For $t\leq 2.0$, $n_T=2$ is seen to be reasonably close to the exact result.
This justifies our choice of $n_T=2$ in Results.
The maximum number of Trotter steps used, $n_T=64$, is well converged to the results from exact diagonalization up to $t=7$.
\begin{figure}[t!]
\centering
  \includegraphics[width=\linewidth]{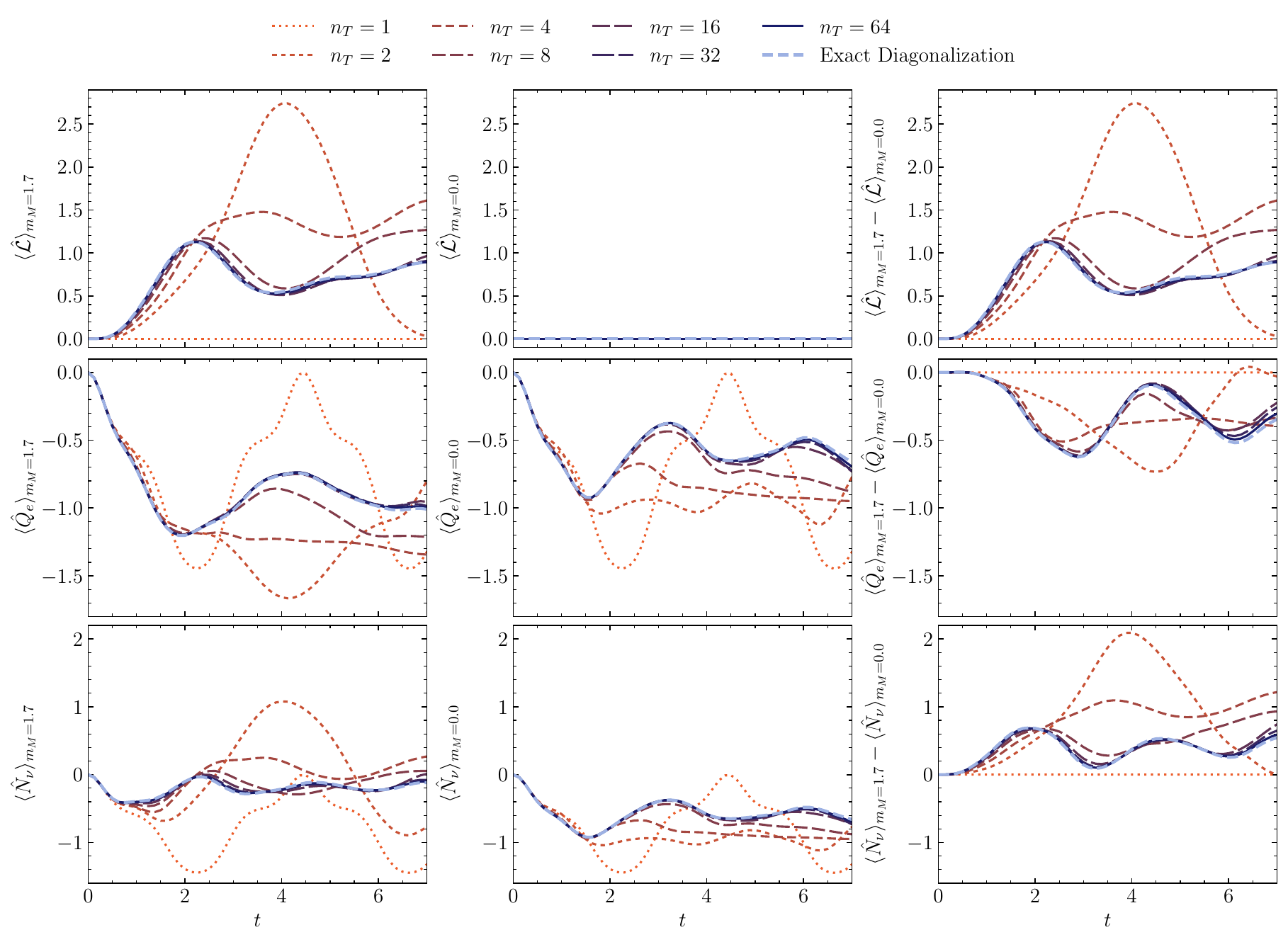}
  \caption{The lepton number, $\hat{\mathcal{L}}$, the electric charge in the lepton sector, $\hat{Q}_e$, and the neutrino number, $\hat{N}_\nu$, computed throughout the time evolution starting from $\vert \psi_{\text{init}}\rangle = |\psi_{\text{vac}}^{(\text{lep})}\rangle|\Delta^- \Delta^-\rangle$
  as a function of the number of Trotter steps.
  These quantities are computed for $L=2$ and for two Majorana masses, $m_M=\{0.0,1.7\}$. No approximations are made in the exact diagonalization results.
  An exact state-vector simulator is used to compute the time-evolution with $n_T$ Trotter steps, which have Trotter errors, as well as (small) errors coming from the SC-ADAPT-VQE preparation of $|\Delta^- \Delta^-\rangle$.
  The quantum simulations that we performed 
  on IonQ's Forte-generation quantum processors  
  employed $n_T=2$ and were limited to 
  $t\le 2$.
  }
  \label{fig:Trotter_results}
\end{figure}

Note that the expected time evolution does not exhibit the well-known 
long-time exponential-decay behavior associated with radioactive decay.
These deviations were discussed in detail in Ref.~\cite{Farrell:2022vyh}, and can be attributed to the 
low-density of final states in a small spatial volume(s).
Simulations of these processes performed in increasing volumes should show convergence to the 
expected exponential behavior.

\subsection{Additional Approximations}
\label{app:circuit_approx}

\begin{figure}[t!]
\centering
  \includegraphics[width=\linewidth]{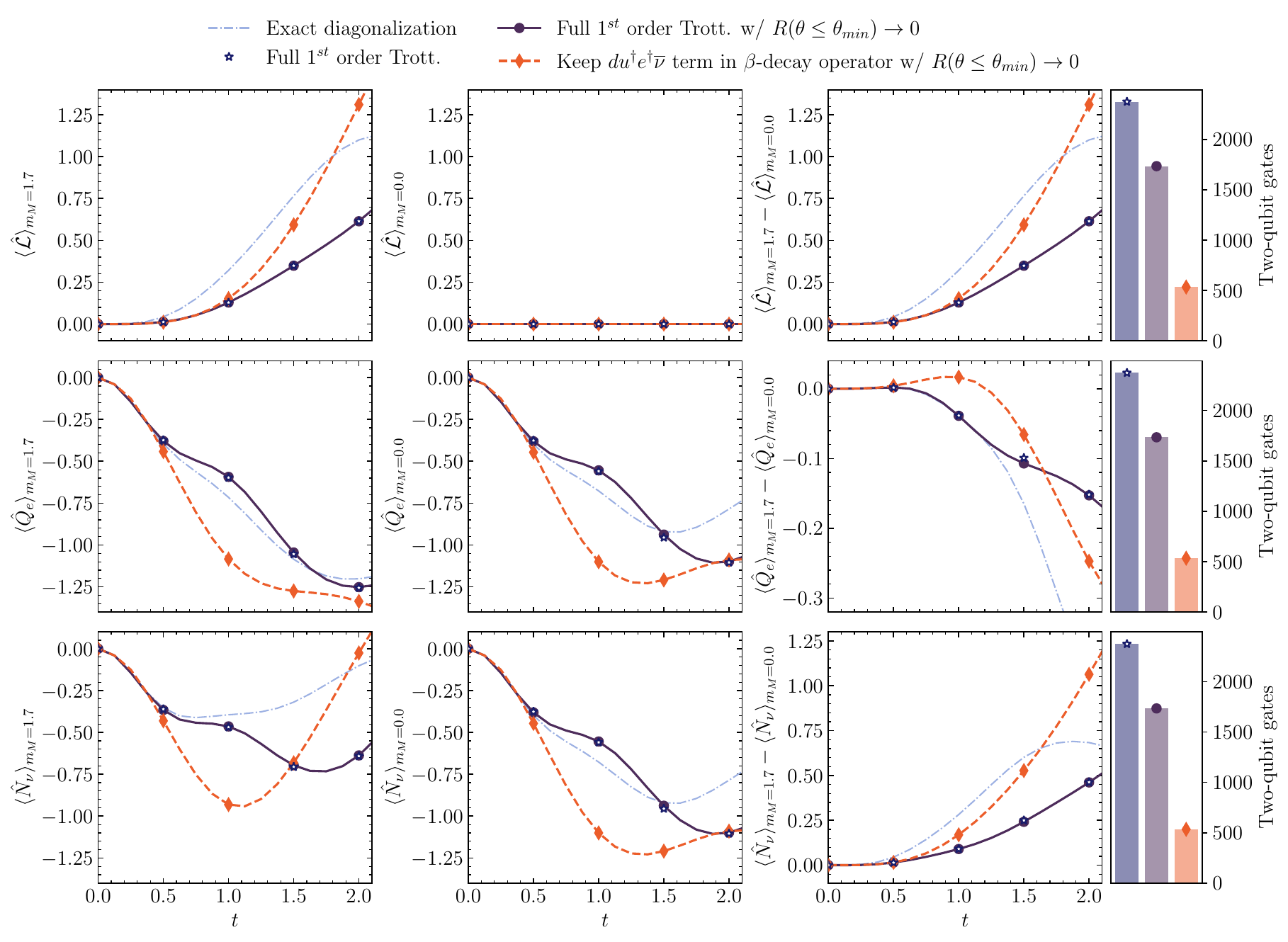}
  \caption{The time-evolution of the lepton number, electric charge and neutrino number with varying levels of approximation, as explained in the text.
  The bar charts to the right give the number of the  $CZ$ and CNOT gates required for each level of approximation. The number of two-qubit gates is the same for both the upper and lower panels, and 
  the approximate time evolution is computed using 
  $n_T=2$ steps of $1^{st}$ order Trotterization.
  The results from full $1^{st}$ order Trotterization  and from 
  the angle-truncated implementation essentially coincide.
  }
  \label{fig:approx}
\end{figure}

\noindent
To study how the approximations 
impact
our simulations, we study several levels of approximations:
\begin{enumerate}[i)]
    \item Exact diagonalization: the initial state preparation and time evolution are performed exactly.
    \item \label{approx:Full}  Full $1^{st}$ order Trotter: $|\Delta^-\Delta^-\rangle$ is approximately prepared with SC-ADAPT-VQE, the range of the chromoelectric interaction is truncated to $\lambda=1$ staggered sites and $n_T=2$ steps of $1^{st}$ order Trotterized time evolution are used.
    \item Full $1^{st}$ order Trotter with $R(\theta\leq \theta_{\text{min}})\to 0$:
    The above approximations plus small rotation angles in the Trotterized time evolution circuits set to zero.
    \item \label{approx:Valence} Keep $du^\dag e^\dag \overline{\nu}$ term: The above approximations plus keeping only the term in $\hat{H}_{\beta}^{1+1}$ that acts on the valence quarks and valence leptons, the first term in \seq~\eqref{eq:BetaHamL2_beta}.
\end{enumerate}
The expectations values of lepton number and lepton electric charge with these approximations 
are shown in \sfig~\ref{fig:approx}.
The bar charts to the right show the number of two-qubit gates (compiled to $CZ$ and CNOT gates) for each one of these approximations.
Approximation~(\ref{approx:Valence})
was used in the experiments performed on IonQ Forte Enterprise, and the IonQ circuit compiler reduced the two-qubit gate count from 534 to 470. In particular, circuits were optimized from 534 down to 454 two-qubit gates, and then inflated to 470 after adding mid-circuit symmetry checks (8 leakage checks were added, totaling 16 extra two-qubit gates).
For the experiments run on IonQ Forte, approximation~(\ref{approx:Full}) was used, and the IonQ circuit compiler reduced the two-qubit gate count from from 2,374 to 2,356 (2,292 two-qubit gates from the time-evolution circuit, plus 64 additional gates from 4 iSWAP checks).
The pre-compilation and post-compilation two-qubit gate counts for 
approximation~(\ref{approx:Full}) and (\ref{approx:Valence}) are given \stab~\ref{tab:gate_counts}.

\begin{table*}[!t]
    \begin{ruledtabular}\begin{tabular}{cccc} 
        Approximation type & Pre-compilation & Post-compilation & Post-compilation + symmetry checks  \\
        \midrule
        (\ref{approx:Valence}) & 534 & 454 & 470 \\
        (\ref{approx:Full}) & 2,374 & 2,292 & 2,356
    \end{tabular}\end{ruledtabular}
    \caption{Two-qubit gate counts for circuits executed on IonQ Forte 
    (approximation~(\ref{approx:Full})) and 
    IonQ Forte Enterprise (approximation~(\ref{approx:Valence})). The gate counts from pre-compilation, post-compilation and post-compilation with additional gates for symmetry checks are shown.}
    \label{tab:gate_counts}
\end{table*}

\section{Additional Details about the Simulation Hamiltonian}
\label{app:HamDetails}
\noindent
This work considers 1+1D QCD with periodic boundary conditions (PBCs) and two flavors of quarks, up $u$ and down $d$.
Discretized onto a staggered lattice with $L$ spatial sites, corresponding to $N=2L$ staggered sites, the Kogut-Susskind Hamiltonian is~\cite{Kogut:1974ag, Banks:1975gq},
\begin{equation}
\hat{H}_{\text{KS}} \ = \ \sum_{f=u,d}\left [\frac{1}{2}\sum_{n=0}^{N-1}\left (\phi^{(f)\dagger}_{n} U_n\phi_{n+1}^{(f)} + {\rm h.c.} \right ) \ + \ m_f \sum_{n=0}^{N-1}(-1)^n\phi^{(f)\dagger}_{n}\phi_{n}^{(f)}\right ]\ + \ \frac{g^2}{2}\sum_{n=0}^{N-1}\sum_{a=1}^8 \vert {\bf E}_n^{(a)} \vert^2  \ ,
\label{eq:HKS}
\end{equation}
where $f =\{u,d\}$ labels the quark flavor, $m_f$ is the quark mass and $g$ is the gauge coupling.
The quark field operators $\phi_n^{(f)}$ are in the fundamental representation ${\bf 3}$ of $SU(3)$, and the three color components $\{r,g,b\}$ have been suppressed for simplicity.
The chromoelectric field ${\bf E}_n^{(a)}$ is in the adjoint representation of $SU(3)$ with $a=\{1,2,\ldots,8\}$.
The spatial gauge link in $A_t^{(a)}=0$ (Weyl) gauge is $U_n$.
Staggered sites related by one period are identified, e.g., $\phi_n^{(f)} = \phi_{n+N}^{(f)}$.

Physical states must satisfy Gauss's law, ${\bf E}_n^{(a)} - {\bf E}_{n-1}^{(a)} = Q_n^{(a)}$, where $Q_n^{(a)} = \sum_f\phi^{(f)\dagger}T^{(a)}\phi^{(f)}$ are the $SU(3)$ charges and the $T^{(a)}$ are the generators of $SU(3)$ in the {\bf 3} representation, i.e., they are $3\times3$ matrices.
With open boundary conditions, these constraints uniquely determine the state of the the gauge field given the state of the fermions.
As a result, explicit gauge field degrees of freedom can be removed, leaving a system of fermions interacting through a Coulomb interaction.
With PBCs, there is one spatial mode of the gauge field that is not constrained.
In order to keep the translational symmetry manifest, it is beneficial to isolate the $k=0$ (zero mode) component of the gauge field~\cite{Dempsey:2022nys},
\begin{align}
\hat{H}_{\text{KS}} \ \to \ &\sum_{f=u,d}\left [\frac{1}{2}\sum_{n=0}^{N-1}\left (\phi^{(f)\dagger}_{n} U \, \phi_{n+1}^{(f)} + {\rm h.c.} \right ) \ + \ m_f \sum_{n=0}^{N-1}(-1)^n\phi^{(f)\dagger}_{n}\phi_{n}^{(f)}\right ]  \nonumber \\
&+ \ \frac{Ng^2}{2}\sum_{a=1}^8\left (E^{(a)}\right )^2 \ + \
\frac{g^2}{2}
\sum_{s=1}^{L}\left (-s + \frac{s^2}{N} \right ) (1 - \frac{1}{2} \delta_{s,L}) \left(\ \sum_{n=0}^{N-1}\sum_{a=1}^8 
Q_n^{(a)} Q_{n+s}^{(a)} \ \right)  \ .
\label{eq:HKSPBC}
\end{align}
Here, $E^{(a)} = \sum_{n=0}^{N-1} E_n^{(a)}/N$ is the zero mode of the electric field, and $U^N = \prod_n U_n$ is its conjugate variable.
The gauge field populates a bosonic Hilbert space that is formally infinite dimensional.
However, it is expected that observables will converge rapidly in the chromoelectric field basis if an increasing number of states are kept.
In this paper, the gauge-field dynamics are frozen, leaving only fermionic degrees of freedom.
The convergence of observables with increasing number of gauge-field states will be quantified in future work.
The Hamiltonian 
that we implement
only has fermionic operators,
\begin{align}
\hat{H}_{\text{KS}} \ \to \ &\sum_{f=u,d}\left [\frac{1}{2}\sum_{n=0}^{N-1}\left (\phi^{(f)\dagger}_{n} \phi_{n+1}^{(f)} + {\rm h.c.} \right ) \ + \ m_f \sum_{n=0}^{N-1}(-1)^n\phi^{(f)\dagger}_{n}\phi_{n}^{(f)}\right ]  \nonumber \\
&
\qquad
+ \
\frac{g^2}{2}
\sum_{s=1}^{L}\left (-s + \frac{s^2}{N} \right )(1-\frac{1}{2}\delta_{s,L}) \left(\ \sum_{n=0}^{N-1}\sum_{a=1}^8 
Q_n^{(a)} Q_{n+s}^{(a)} \ \right)  \ .
\label{eq:HKSPBCv2}
\end{align}
Similar Hamiltonians that have used Gauss's law to remove gauge degrees of freedom in 1+1D lattice gauge theories with PBCs were used in Refs.~\cite{Klco:2018kyo,Zache:2018cqq,Nagele:2018egu,Davoudi:2025rdv,Chai:2025qhf,Farrell:2025nkx}.

\subsection{Truncating the Chromoelectric Interaction}
\label{sec:truncHam}
\noindent
In theories with a mass gap between the vacuum and first excited state, like 1+1D QCD, correlations in the vacuum decay exponentially between distant charges,
\begin{equation}
\sum_{a=1}^8\langle Q_n^{(a)} Q_{n+s}^{(a)}\rangle \ \sim \ e^{-c_1 s \, m_{\text{hadron}}} \ ,
\end{equation}
where $c_1$ is a geometrical factor, and $m_{\text{hadron}}$ is the mass of the lightest excitation in the vacuum sector, the scalar meson. 
Vacuum expectation values are denoted by $\langle \, . \, \rangle$.
In previous work~\cite{Farrell:2024fit}, the exponential decay of correlations between electric charges in the Schwinger model (one-dimensional $U(1)$ lattice gauge theory) was used to form an approximate interaction where charges interacting beyond a certain range were truncated.\footnote{In $U(1)$ gauge theory, due to staggering, it is only the connected part of this correlation that falls exponentially.
In $SU(3)$ gauge theory, the disconnected part of the correlation vanishes when evaluated in physical color singlet states.}
A similar strategy is used here to truncate interactions between color charges separated by more than $\lambda$ staggered sites as,
\begin{equation}
\hat{H}_{el}(\lambda) \ = \ \frac{g^2}{2}
\sum_{s=1}^{\lambda}\left (-s + \frac{s^2}{2L} \right )(1-\frac{1}{2} \delta_{s,L}) \left(\ \sum_{n=0}^{2L-1}\sum_{a=1}^8 
Q_n^{(a)} Q_{n+s}^{(a)} \right ) \ .
\label{eq:Hellambda}
\end{equation}
This truncated interaction reduces the two-qubit gate count required for time evolution from scaling as $\mathcal{O}(N^2)$ to $\mathcal{O}(\lambda N)$.
The quantum simulations performed on IonQ Forte-generation QPUs in Results used $\lambda=1$.

\subsection{The Weak Interaction}
\label{sec:weakHami}
\noindent
The weak interactions giving rise to single-$\beta$ decay are modeled through a local vector-like four-Fermi operator~\cite{Farrell:2022vyh},
\begin{eqnarray}
    \hat{H}_\beta^{1+1} & = & 
    \frac{G}{\sqrt{2}} \int\! d^2 x \left (\overline{\psi}_u\gamma^\mu \psi_d\ 
    \overline{\psi}_e\gamma_\mu \mathcal{C} \psi_{\nu}  
         + {\rm h.c.} \right )
    \ \nonumber \\
    & \to &
    \frac{G}{ \sqrt{2}} 
    \sum_{n \ {\rm even}} 
    \bigg [
    \left (\phi_{n}^{(u)\dagger} \phi_{n}^{(d)} + \phi_{n+1}^{(u)\dagger} \phi_{n+1}^{(d)} \right ) \left (\chi_{n}^{(e)\dagger} \chi_{n+1}^{(\nu)} - \chi_{n+1}^{(e)\dagger} \chi_{n}^{(\nu)}\right ) 
    \nonumber\\
 && 
 \qquad 
 + \left ( \phi_{n}^{(u)\dagger} \phi_{n+1}^{(d)} + \phi_{n+1}^{(u)\dagger} \phi_{n}^{(d)} \right )
    \left (\chi_{n}^{(e)\dagger} \chi_{n}^{(\nu)} - \chi_{n+1}^{(e)\dagger} \chi_{n+1}^{(\nu)}\right )+
    {\rm h.c.} 
    \bigg ] 
    \nonumber \\
    &\approx &
    \frac{G}{\sqrt{2}} \  \sum_{n \ \rm{even}}\left (\phi_n^{(u)\dagger}\phi_n^{(d)}\phi_n^{(e)\dagger}\phi_{n+1}^{(\nu)}
         + {\rm h.c.} \right ) 
         \ ,
    \label{eq:HbetaC1}
\end{eqnarray}
where $\mathcal{C}$ is the charge-conjugation operator, $\gamma^{\mu}$ are the gamma-matrices and $G$ is the weak coupling constant (Fermi's constant).
The first line shows the 1+1D interaction related to the low-energy charged-current weak interaction of the Standard Model.
The mapping of the fermion fields to a staggered lattice is shown in the second line.
This necessarily includes contribution from both particles and anti-particles due to operator contractions.
The results displayed in Fig.~2 in the main text are obtained using this interaction.
The third line employs a ``valence-fermion" approximation that only keeps the terms acting on the valence-quarks (no operators acting on anti-quark sites) and valence-leptons (no operators acting on neutrino or anti-electron sites).
The results displayed in Fig.~3 in the main text are obtained using this approximation.

\subsection{The Complete Spin Hamiltonian}
\label{app:fullspinhami}
\noindent
The quantum simulations described in Results used $L=2$ spatial sites and the truncated chromoelectric interaction in \seq~\eqref{eq:Hellambda} with $\lambda=1$. The ordering of the fermionic degrees of freedom is shown in Fig.~4 in the main text. 
The JW transformation is used to map fermionic to spin operators that acts on qubits, 
\begin{equation}
\phi^{\dagger}_i =  \prod_{j < i} \left(-\hat{Z}_j\right) \hat{\sigma}^{+}_i 
\quad ,\quad 
\phi_i \ =\  \prod_{j < i} \left(-\hat{Z}_j\right) \hat{\sigma}^{-}_i 
\ ,
\label{JWmapdefop}
\end{equation}
where $\phi^{\dagger}$ and $\phi$ are fermionic creation and annihilation operators, $\hat{Z}$ is the Pauli-Z operator, and $\hat{\sigma}^{+}$ and $\hat{\sigma}^{-}$ are the spin-raising and spin-lowering operators.

After the JW transformation is applied, the complete spin Hamiltonian is
\begin{subequations}
\label{eq:BetaHamL2}
\begin{align}
    \hat{H}_{\rm{quarks}} \rightarrow &\
    \frac{1}{2} \sum_{n=0}^{3} \sum_{f=0}^1\sum_{c=0}^{2}  m_f\left [\left (-1\right )^n \hat{Z}_{6n+3f+c} +\hat{I}\right ] \nonumber \\
    & -\frac{1}{2} \sum_{n=0}^{2} \sum_{f=0}^1 \sum_{c=0}^{2}\left [\hat{\sigma}^+_{6n+3f+c}\hat{Z}^5\hat{\sigma}^-_{6n+6+3f+c}+{\rm h.c.}\right ] + \frac{1}{2}\sum_{f=0}^1 \sum_{c=0}^{2}\left [\hat{\sigma}^+_{18+3f+c}\hat{Z}^5\hat{\sigma}^-_{3f+c}+{\rm h.c.}\right ]\ , \label{eq:BetaHamL2_quarkkin} \\[4pt]
    \hat{H}_{\rm{leptons}} \rightarrow  
    &\ \frac{1}{2} \sum_{n=0}^{3} \sum_{f=0}^1 m_f\left [\left (-1\right )^n \hat{Z}_{24+2n+f} +\hat{I}\right ] 
    + \frac{1}{2} \sum_{l = 0}^{1}m_M\left[ 
    \hat{\sigma}^+_{24+4l}\ \hat{Z} \hat{\sigma}^+_{26+4l} \: + \: {\rm h.c.} \right] \nonumber \\
    &-\frac{1}{2} \sum_{n=0}^{2} \sum_{f=0}^1 \left [\hat{\sigma}^+_{24+2n+f}\hat{Z}\hat{\sigma}^-_{26+2n+f}+{\rm h.c.}\right ] + \frac{1}{2}\sum_{f=0}^1 \left [\hat{\sigma}^+_{30+f}\hat{Z}\hat{\sigma}^-_{24+f}+{\rm h.c.}\right ] \ , \label{eq:BetaHamL2_lepkin} \\[4pt]
    \hat{H}_{{\rm glue}}
    \rightarrow  & - \frac{g^2}{2}\sum_{n=0}^{3} \frac{3}{4}\hat{Q}_{n}^{(a)} \hat{Q}_{n+1}^{(a)} \ , \label{eq:BetaHamL2_glue} \\[4pt]
    \hat{H}_{\beta}^{1+1} \rightarrow 
    & \ \frac{G}{\sqrt{2}}\sum_{l = 0}^{1}\sum_{c=0}^2 \bigg [
    \hat{\sigma}^-_{24+4l+2}\hat{\sigma}^+_{24+4l+1} \hat{\sigma}^-_{12l+3+c} \hat{Z}^2 \hat{\sigma}^+_{12l+c} \: - \: \hat{\sigma}^+_{24+4l+3} \hat{Z}^2 \hat{\sigma}^-_{24+4l} \hat{\sigma}^-_{12l+9+c} \hat{Z}^2 \hat{\sigma}^+_{12l+6+c}  \nonumber \\[4pt]
    & -  \hat{\sigma}^+_{24+4l+3}\hat{Z}^2\hat{\sigma}^-_{24+4l} \hat{\sigma}^-_{12l+3+c} \hat{Z}^2 \hat{\sigma}^+_{12l+c}   \: + \: \hat{\sigma}^-_{24+4l+2} \hat{\sigma}^+_{24+4l+1} \hat{\sigma}^-_{12l+9+c} \hat{Z}^2 \hat{\sigma}^+_{12l+6+c} \nonumber \\[4pt] 
    & + \: \hat{\sigma}^+_{24+4l+1} \hat{\sigma}^-_{24+4l} \hat{\sigma}^+_{12l+6+c} \hat{Z}^2 \hat{\sigma}^-_{12l+3+c} \: - \: \hat{\sigma}^+_{24+4l+3} \hat{\sigma}^-_{24+4l+2} \hat{\sigma}^-_{12l+9+c}\hat{Z}^8 \hat{\sigma}^+_{12l+c} \nonumber \\[4pt]
     &+ \: \hat{\sigma}^+_{24+4l+1} \hat{\sigma}^-_{24+4l} \hat{\sigma}^-_{12l+9+c} \hat{Z}^8 \hat{\sigma}^+_{12l+c} \: - \: \hat{\sigma}^+_{24+4l+3} \hat{\sigma}^-_{24+4l+2} \hat{\sigma}^+_{12l+6+c} \hat{Z}^2 \hat{\sigma}^-_{12l+3+c} \: + \: {\rm h.c.} \bigg] \ . \label{eq:BetaHamL2_beta}
\end{align}
\end{subequations}
Here, $\hat{H}_{\text{quarks}}+\hat{H}_{\text{leptons}} = \hat{H}_{\text{free}}$ (defined in Eq.~3 in the main text) and the hopping term between the last and first staggered site has a relative minus sign due to the JW mapping. 
The index $l$ in $\hat{H}^{1+1}_{\beta}$ labels the spatial sites.
We have used the short-hand notation $\hat{\sigma}_i\hat{Z}^{(j-i-1)}\hat{\sigma}_j=\hat{\sigma}_i(\prod_{k=i+1}^{j-1}\hat{Z}_k)\hat{\sigma}_j$.
In $\hat{H}_{\text{quarks}}$, $m_0=m_u$ and $m_1=m_d$.
In $\hat{H}_{\text{leptons}}$, $m_0=m_\nu$ and $m_1=m_e$.
The product of $SU(3)$ charges are
\begin{align}
    \sum_{a=1}^8 \hat{Q}_{n}^{(a)} \, \hat{Q}_{m}^{(a)} = 
    \sum_{f=0}^1\sum_{f'=0}^1
    \ \frac{1}{4} &\Bigl[\ 
    2\left(  \hat{\sigma}^+_{6n+3f}\hat{\sigma}^-_{6n+3f+1}\hat{\sigma}^-_{6m+3f'}\hat{\sigma}^+_{6m+3f'+1} 
    \right. 
    \nonumber\\
    & \left. \qquad 
    +\ \hat{\sigma}^+_{6n+3f}\hat{Z}_{6n+3f+1}\hat{\sigma}^-_{6n+3f+2}\hat{\sigma}^-_{6m+3f'}\hat{Z}_{6m+3f'+1}\hat{\sigma}^+_{6m+3f'+2} 
    \right.  
    \nonumber\\
    & \left. \qquad
    +\ \hat{\sigma}^+_{6n+3f+1}\hat{\sigma}^-_{6n+3f+2}\hat{\sigma}^-_{6m+3f'+1}\hat{\sigma}^+_{6m+3f'+2} + { \rm h.c.} \right)   
    \nonumber\\
    &
    +\ \frac{1}{6}\sum_{c=0}^{2} \sum_{c'=0}^2( 3 \delta_{c c'} - 1 ) \hat{Z}_{6n+3f+c}\hat{Z}_{6m+3f'+c'} 
    \ \Bigr] 
    \ .
    \label{eq:QnfQmfp}
\end{align}
For completeness, we provide the total electric charge $\hat{Q}_{\text{tot}}$ (quark $+$ lepton) and the diagonal color charges (redness $\hat{r}$, blueness $\hat{b}$ and greenness $\hat{g}$),
\begin{align}
\hat{r} \ = \ &\frac{1}{2}\sum_{n=0}^3\sum_{f=0}^1  \hat{Z}_{6n + 3f}  \ , \nonumber \\
\hat{g} \ = \ &\frac{1}{2}\sum_{n=0}^3\sum_{f=0}^1\hat{Z}_{6n + 1 + 3f} \ , \nonumber \\
\hat{b} \ = \ &\frac{1}{2}\sum_{n=0}^3\sum_{f=0}^1 \hat{Z}_{6n + 2 + 3f} \ , \nonumber \\
\hat{Q}_{\text{tot}} \ = \  &\frac{1}{2}\sum_{n=0}^3\sum_{f=0}^1\sum_{c=0}^2 q_f \hat{Z}_{6n + 3f + c}  \ - \ \frac{1}{2}\sum_{n=0}^3 \hat{Z}_{25+2n}  \ ,
\end{align}
where $q_0 = 2/3$ is the electric charge of the up quark and $q_1 = -1/3$ is the electric charge of the down quark.
The initial state $|\Delta^- \Delta^-\rangle$ states used in our quantum simulations has $r=g=b=2$ and $Q_{\text{tot}}=-2$.
These charges are conserved under time evolution, but can be violated by device errors.
The results obtained from IonQ Forte Enterprise were post-selected to conserve these charges.

After the JW mapping, the complete charged-current weak interaction (modeled in 1+1D) is given in \seq~\eqref{eq:BetaHamL2_beta}.
It includes operators acting on both the valence- and sea-fermions, i.e., an up-quark operator can create an up-quark or it can annihilate an anti-up-quark, both actions change the up-quark number by one.  For a complete simulation, all possible such actions should be included.  
However, retaining only the contributions from the operators acting on valence quarks is a well-defined approximation that can be implemented, reducing the required depth of quantum circuits.  
\seq~\eqref{eq:ValenceJWweak} gives the Hamiltonian describing only the valence weak interactions, the approximation that we make in obtaining the results shown in Fig.~3 and shown in Table~2 in the main text,
\begin{equation}
\hat{H}_{\beta, {\rm valence}}^{1+1} =
    \frac{G}{\sqrt{2}} \sum_{l = 0}^{1}\sum_{c=0}^2 
\     \hat{\sigma}^-_{24+4l+2}\hat{\sigma}^+_{24+4l+1} \hat{\sigma}^-_{12l+3+c} \hat{Z}^2 \hat{\sigma}^+_{12l+c} 
\ \ .
\label{eq:ValenceJWweak}
\end{equation}
%

\section{Bootstrapping for Uncertainty Estimation}
\label{app:bootstrapping}
\noindent
As the outputs of the quantum simulation we perform are computed with both post-selection and the non-linear filtering, estimating the uncertainty of the estimated outputs is non-trivial. We perform a bootstrap~\cite{Bootstrap_Efron_1979} estimation of these error bars based on the two sources of variation: the choice of circuit variants and the individual shots for each of those variants. For $N_V$ variants and $n_s$ shots per variant, we re-sample our measurements by taking the following steps:
\begin{enumerate}
    \item Sample, with replacement, $N_V/2$ pairs of variants among the set of variant pairs used. (Recall from Methods~C that the variants are generated in pairs, with each pair differing only by bit flips before measurement to symmetrize readout errors.)
    \item Combine all pairs selected in the previous step into a set of $N_V$ variants. For each of the resampled variants, sample (with replacement) $n_s$ bit strings from the histograms of bit strings for each variant.
    \item Compute the simulation outputs for the resampled data, using the same post-selection and non-linear filtering described above.
\end{enumerate}

By repeating this resampling procedure we construct a population of resampled outputs, and this population is used to estimate the standard-deviation of those outputs. For this work we used $10^5$ bootstrap samples to estimate these standard-deviations. We then use these standard deviations as our reported error bars. 

We note that this bootstrapping approach has a weakness: it does not account for the intentional structure of the choice of variants. Since the variants are chosen to have different biases due to noise in such a way that they should cancel, modeling them as random draws over-estimates the variance due to this selection. However, as modeling the precise uncertainty due to this bias cancellation would require an impractically detailed and accurate description of the device noise (including drift in that noise over time), we choose to use the random-draw model for bootstrapping and err on the side of over-estimating the uncertainties.

\section{Error Mitigation with Flag Gadgets}
\label{app:em_flag}
\noindent
Flag gadgets can be used to verify mid-circuit symmetries derived from the commutation properties of subcircuit blocks~\cite{Ken2020}. In quantum chemistry, as in QCD, one of the most common subcircuit blocks expresses particle-preserving unitaries, an extension of Givens rotations~\cite{Arrazola2022}, which have multiple SWAP symmetries. For example, four-qubit Givens~\cite{Goings2023} rotations, which are equivalent to four-Fermi operators $\theta \sigma^-_i \sigma^+_j \sigma^-_k \sigma^+_l + {\rm h.c.}$ up to a phase factor,\footnote{Givens rotations (with real matrix elements) express unitary transformations by  the exponential of $\theta \sigma^-_i \sigma^+_j \sigma^-_k \sigma^+_l - {\rm h.c.}$ (note the minus sign).
} 
commute with SWAP gates applied on $i-k$ or $j-l$ qubit pairs and negated SWAP gates applied on $i-j$ or $k-l$ qubit pairs. 
Since four-Fermi operators also commute with $R_{ZZ}$ gates, it means they also commute with iSWAP and negated iSWAP gates, as shown in \sfig~\ref{fig:symmetry_givens}.

We chose iSWAP symmetry checks over $R_{ZZ}$ and SWAP checks because of IonQ Forte and Forte Enterprise having low  X- and Y-errors from crosstalk or under-rotation compared to phase errors. Optimal construction of symmetry checks such as iSWAP checks or negated iSWAP checks, which naively require 16 two-qubit gates, can be reduced to only 6 additional two-qubit gates per check (see \sfig~\ref{fig:symmetry_check}). We group together $\hat{H}^{1+1}_{\beta}$ terms applied on the same lepton qubits (light-brown circuit blocks in  \sfig~\ref{fig:BetaCirc}) and apply negated iSWAP checks around them to ensure that the number of added gates (6 $R_{ZZ}$ gates) is less than the number of gates in the checked subcircuit (33 $R_{ZZ}$ gates). All iSWAP checks are applied on the lepton register. Used together with dynamical decoupling, mid-circuit symmetry checks improve the circuit fidelity upon post-selection, as shown in \stab~\ref{tab:em_methods}.
\begin{figure}[t]
    \centering
   \begin{tikzpicture}
        \node[scale=1] {
            \begin{quantikz}[row sep={0.9cm,between origins}]
                \lstick{$\ket{a_0}$} & \gate{H} & \ctrl{4} & \ctrl{4} & & & & & & & & \ctrl{4} & \ctrl{4} & \gate{H} & \\
                \lstick{$\ket{q_i}$} & & & & \ctrl{3}\gategroup[4,steps=7,style={inner sep=1pt, color=orange}, label style={label position=below, yshift=-0.5cm}]{} & \ctrl{2} & \ctrl{1} & \gate{R_X(\theta)} & \ctrl{1} & \ctrl{2} & \ctrl{3} & & &  &  \\
                \lstick{$\ket{q_j}$} & & & & & & \targ{} & \ctrl{-1} & \targ{} & & & & & & \\
                \lstick{$\ket{q_k}$} & & \targ{} & \gate[2]{i\textsc{SWAP}} & & \targ{} & & \ctrl[open]{-1} & & \targ{} & & \gate[2]{i\textsc{SWAP}^\dagger} & \targ{} & & \\
                \lstick{$\ket{q_l}$} & & \targ{} &  & \targ{} & & & \ctrl{-1}& & & \targ{} &  & \targ{} & & 
            \end{quantikz}
        };
    \end{tikzpicture}
    \caption{Negated iSWAP-check placement around the unitary generated by the four-Fermi operator, 
    $\theta \sigma^-_i \sigma^+_j \sigma^-_k \sigma^+_l+ {\rm h.c.}$ (the orange box),  on $k-l$ qubits. Controlled iSWAP with two controlled NOTs commute with the four-Fermi operator and do not change the state of the ancilla $\left|a_0\right>$ in the absence of errors.
    }
  \label{fig:symmetry_givens}
\end{figure}

\begin{figure}[t]
    \centering
   \begin{tikzpicture}
        \node[scale=1] {
            \begin{quantikz}[row sep={0.9cm,between origins}]
                \lstick{$\ket{a_0}$} &  &  &  & \ctrl{1} &  &  &  & \ctrl{1} &  &  &  &  &  \\
                \lstick{$\ket{q_k}$} & \ctrl{1} & \gate{H} & \gate[style={color=orange}]{T^\dagger / T} & \targ{} & \gate[style={color=orange}]{T / T^\dagger} & \targ{} & \gate{T^\dagger} & \targ{} & \gate{T}  & \gate{H} & \ctrl{1} & \gate{S^\dagger} &  \\
                \lstick{$\ket{q_l}$} & \targ{} &  &  &  &  & \ctrl{-1} & \gate{S^\dagger} &  &  &  & \targ{} & \gate{S} &
            \end{quantikz}
        };
    \end{tikzpicture}
    \caption{Optimized construction of a controlled iSWAP (or a controlled negated iSWAP if the marked $T$ gates are inverted) requires only five two-qubit gates. When placed on each side of the optimized Givens rotations, $k-l$ CNOTs cancel out leading to only 6 two-qubit gates overhead per check.
}
  \label{fig:symmetry_check}
\end{figure}

Leakage and qubit loss errors in quantum computers lead to qubits becoming unresponsive to quantum gates. Undetected leakage errors can severely corrupt quantum computation, especially if they happen early in the circuit. In ion-trapped devices, leakage errors can happen due to the spontaneous emission and electronic transitions outside of the computational space.
Some known solutions use custom-designed gates that target the extended subspace to which the leakage occurs~\cite{Leakage2024}.
We make use of a different approach with a combination of two fully entangling gates and single-qubit flips to flag leakage events when entangling fails~\cite{Baldwin2025,Stricker2020}. Our construction of this check is shown in \sfig~\ref{fig:leakage_check}.

\begin{figure}[t!]
    \centering
   \begin{tikzpicture}
        \node[scale=1] {
            \begin{quantikz}[column sep=1cm]
                \lstick{$\ket{1}$} & \gate{R_Y(\pi/2)} & \gate[2]{R_{ZZ}(\pi/2)}  &  & \gate[2]{R_{ZZ}(-\pi/2)} & \gate{R_Y(-\pi/2)} &  \\
                \lstick{$\ket{\psi}$} &  &  & \gate{Y} &  & \gate{X} &  
            \end{quantikz}
        };
    \end{tikzpicture}
   \caption{The ancilla qubit (shown as the top wire) is a flag qubit that is prepared in state $\left|1\right>$ (or $\left|0\right>$). The target qubit on which the leakage is tested can be in any state $\left|\psi \right>$ since the following sequence of gates does not affect it. The ancillary qubit ends up in state $\left|0\right>$ (or $\left|1\right>$) 
   if the target qubit is 
   in the computational space.
   }
  \label{fig:leakage_check}
\end{figure}

\begin{table*}[!t]
    \begin{ruledtabular}\begin{tabular}{llcccc} 
        $t$ & \makecell{Error Mitigation} & $\langle\hat{\mathcal{L}}\rangle_{m_M=0}$ & $\langle\hat{\mathcal{L}}\rangle_{m_M=1.7}$& $\langle\hat{Q}_e\rangle_{m_M=0}$ & $\langle\hat{Q}_e\rangle_{m_M=1.7}$  \\
        \midrule
        \multirow{4}{*}{0.5} & No EM & $-0.04 \pm 0.02$ & $-0.05 \pm 0.02$ & $-0.12 \pm 0.01$ & $-0.10 \pm 0.01$\\
        &  DNL & $-0.16 \pm 0.02$ & $-0.14 \pm 0.02$ & $-0.26 \pm 0.02$ & $-0.26\pm0.02$\\
        &  DNL + PS & $-0.06 \pm 0.08$ & $-0.10 \pm 0.07$ & $-0.54 \pm 0.04$ & $-0.52 \pm 0.05$\\
        &  DNL + PS + FG & $-0.06\pm 0.08$ & $-0.09\pm 0.07$ & $-0.52\pm 0.05$ & $-0.54\pm0.05$  \\
        \midrule
        \multirow{4}{*}{1.0} & No EM & $-0.04 \pm 0.02$ & $-0.01 \pm 0.02$ & $-0.17 \pm 0.02$ & $-0.16 \pm 0.01$ \\
        &  DNL & $-0.12 \pm 0.03$ & $-0.01 \pm 0.02$ & $-0.37 \pm 0.02$ & $-0.36 \pm 0.02$\\
        &  DNL + PS & $\phantom{-}0.04 \pm 0.07$  & $\phantom{-}0.11 \pm 0.08$ & $-0.89 \pm 0.06$ & $-0.87\pm0.06$ \\
        &  DNL + PS + FG & $\phantom{-}0.00 \pm 0.06$ & $\phantom{-}0.12\pm0.09$ & $-0.90\pm0.07$ & $-0.86\pm 0.08$  \\
        \midrule
        \multirow{4}{*}{1.5} & No EM & $-0.02 \pm 0.02$ & $\phantom{-}0.07 \pm 0.01$ & $-0.20 \pm 0.02$ & $-0.19 \pm 0.01$ \\
        & DNL & $-0.10 \pm 0.03$ & $\phantom{-}0.19 \pm 0.02$ & $-0.4 \pm 0.02$ & $-0.43 \pm 0.02$\\
        &  DNL + PS & $\phantom{-}0.09 \pm 0.06$ & $\phantom{-}0.68 \pm 0.11$ & $-1.05 \pm 0.06$ & $-1.24 \pm 0.06$\\
        &  DNL + PS + FG & $\phantom{-}0.05\pm0.05$ & $\phantom{-}0.59\pm0.11$ & $-1.04\pm0.06$ & $-1.25\pm 0.07$  \\
        \midrule
        \multirow{4}{*}{2.0} & No EM  & $-0.02 \pm 0.01$ & $\phantom{-}0.28 \pm 0.01$ & $-0.31 \pm 0.01$ & $-0.30 \pm 0.01$\\
        & DNL & $-0.03 \pm 0.02$ & $\phantom{-}0.37 \pm 0.02$ & $-0.45 \pm 0.01$ & $-0.41 \pm 0.01$\\
        &  DNL + PS  & $\phantom{-}0.13 \pm 0.07$ & $\phantom{-}1.41 \pm 0.09$ & $-1.10 \pm 0.05$ & $-1.38 \pm 0.05$\\
        &  DNL + PS + FG & $\phantom{-}0.08\pm 0.07$ & $\phantom{-}1.43\pm 0.12$ & $-1.13\pm 0.05$ & $-1.41\pm 0.06$ \\
    \end{tabular}\end{ruledtabular}
    \caption{Forte experimental results of $\langle\hat{\mathcal{L}}\rangle$ and $\langle\hat{Q}_e\rangle$ for  $m_M=0$ and $m_M=1.7$ at different evolution times $t$. The results are shown for four distinct levels of error mitigation (EM). The ``No EM" data represents the raw experimental output. Applying debiasing with non-linear filtering yields the ``DNL" results. The ``DNL + PS" data further includes post-selection (PS) on the total charge $Q_e$ without applying post-selection based on flag gadgets. Finally, the ``DNL + PS + FG" data shows the fully mitigated results, which adds post-selection on flag gadgets (FG) and corresponds to the data in Table~2 in the main text.}
    \label{tab:em_methods}
\end{table*}

\section{Additional Details on the  Selection of Parameters}
\label{app:spectrum}
\noindent
The following parameters are used in our simulations
(in lattice units),
\begin{align}
m_u = 1 \ , \ m_d  = 1.5 \ , \ m_e = 0.1 \ , \  m_{\nu} = 1.5 \ , \ m_M = \{0.0, 1.7\} \ , \ g=1 \ , \ G=1 \ , \ L=2 \ .
\label{eq:params}
\end{align}
The corresponding spectra obtained from exact diagonalization using these parameters are given in \stab~\ref{tab:spec}.
The energies are calculated from the Hamiltonian without weak interactions, i.e., $\hat{H} = \hat{H}_{\text{free}} + \hat{H}_{\text{glue}}+ \hat{H}_{\text{Maj}}$.
The left 
columns in \stab~\ref{tab:spec} show
 that, for $m_M=0$, the states after a $\beta$- and $\beta\beta$-decay are higher in energy than the initial state.
The right columns show
that, for $m_M=1.7$, only the state after single $\beta$-decay is higher in energy; the state after $0\nu\beta\beta$ is lower in energy.\footnote{
The lepton vacuum for $m_M=1.7$ and $m_M=0$ are identical, with ${\cal L}=0$.
In general, for $m_M \neq 0$, ${\cal L}$ is not a good quantum number and, for example, the state labeled by $\overline{\nu}$ in the second row of the right side of \stab~\ref{tab:spec} has components with ${\cal L}=-1$ and ${\cal L}=+1$.
}
Therefore, these parameters engineer an energy landscape that is similar to that found 
in, for example, $^{76}\text{Ge}$, 
i.e., states obtained via a single $\beta$-decay from $|\Delta^-\Delta^-\rangle$ are higher in energy.
However, these are the energies in the absence of the $\beta$-decay interaction.
Due to the size of the Fermi constant ($G=1$), as well as finite-size effects, decays are still allowed for $m_M=0$, even though they are energetically disfavored.
Of course, with $m_M=0$, lepton number is conserved, and neutrinoless decays cannot occur.
\begin{table}[!ht]
\renewcommand{\arraystretch}{1.2}
\begin{tabularx}{0.4\textwidth}{l Y} 
\hline \hline
\multicolumn{2}{c}{$m_M = 0$} \\
 \hline
 \multicolumn{1}{c}{State} & Energy\\
 \hline 
 $\vert \Delta^-\Delta^-\rangle$ & 6.7332\\ 
 $\vert \Delta^-\Delta^0\rangle+e^-+\overline{\nu}$ & 7.7241\\ 
 $\vert \Delta^0\Delta^0\rangle+2e^-+2\overline{\nu}$ & 9.9384\\ 
 \hline \hline
\end{tabularx}
\quad
\begin{tabularx}{0.4\textwidth}{l Y} 
\hline \hline
\multicolumn{2}{c}{$m_M = 1.7$} \\
 \hline
 \multicolumn{1}{c}{State} & Energy\\
 \hline 
 $\vert \Delta^-\Delta^-\rangle$ & 6.7332\\ 
 $\vert \Delta^-\Delta^0\rangle +e^-+\overline{\nu}$ & 6.8742\\ 
 $\vert \Delta^0\Delta^0\rangle+2e^-$ & 6.6356\\ 
 \hline \hline
\end{tabularx}
\caption{
The energies of states (in lattice units)
relevant to $\beta$-decays, $2\nu\beta\beta$-decays and $0\nu\beta\beta$-decays, with the parameters given in \seq~\eqref{eq:params}.}
\label{tab:spec}
\end{table}
%

\section{Comments on the Road Ahead for Simulation Parameter Extrapolation(s)}
\label{sec:paramextraps}
\noindent
The road map for quantum simulations of fundamental physics is anticipated to mirror that of classical lattice QCD calculations.
Starting in the mid-1970's, lattice QCD simulations were performed using selection of truncations and unphysical parameters, including finite-volume, unphysical quark masses, finite lattice spacings, quenching (valence quark contributions only) 
followed by partial-quenching 
(valence and sea-quarks included, but with different Hamiltonians).
As computational capabilities increased, lattice QCD simulations were increasingly able to be extrapolated to the physical point with finite-volume and continuum extrapolations. 
For a review of uncertainty quantification in lattice QCD calculations, see Ref.~\cite{Beane:2014oea}.

The current status of quantum simulations should be considered to be in the era of the 1970's lattice QCD simulations, but with the additional overhead of quantum computers 
producing observable dependent errors in simulations. 
Even after we evolve to the point of present day simulations, 
extrapolations in the parameters of the neutrino sector, from the simulation values to their physical (or limit values),
will be required.\footnote{Quantum simulations of neutrinos~\cite{Hall:2021rbv,Yeter-Aydeniz:2021olz,Illa:2022jqb,Amitrano:2022yyn,Illa:2022zgu,Siwach:2023wzy,Turro:2024shh,Spagnoli:2025etu} are also being used to explore collective flavor dynamics that occur in high-density non-equilibrium environments, such as those formed in supernova.}  Given the extreme hierarchy of scales, such extrapolations are expected to be straightforward once simulations have been performed with sufficient fidelity.

A further added complication for the quantum simulations is the extraction of lifetimes from simulations.  The time evolution of the change in final-state lepton number becomes exponential for large volumes and at late times.   However, before these conditions are satisfied, the exponential decay is only approximate, from which the mean-lifetime must be extracted.  This was studied in detail in 1+1D for $\beta$-decay in Ref.~\cite{Farrell:2022vyh}.

\section{Debiasing with Non-linear Filtering}
\label{app:em_dnl}
\noindent
Debiasing with non-linear filtering (DNL) is a resource-efficient error-mitigation strategy that leverages circuit and device symmetries~\cite{symm2023}. 
In this approach we prepare a number of variants of a quantum circuit that correspond to the same ideal computation but differ in the embedding of the algorithmic qubits to ions as well as gate decomposition and implementations. By choosing these variants to symmetrize the effects of the dominant hardware errors, the effect of those errors can be reduced in post processing by applying a filtering function.

This debiasing method can be highly efficient as these variants are carefully selected as opposed to being random draws (such as is done with standard twirling).
For example, given a measurement bias between 
the $\ket{0}$ and $\ket{1}$ states
in the hardware, one could always prepare an even number of variants grouped into pairs that are identical up to attaching NOT gates at the end of the circuit to the corresponding qubits.
Undoing those NOT operations in postprocessing will then cancel out the measurement bias for each pair, assuming there is no significant drift in that bias while the pair of variants are collected. Another relevant symmetry available on the Forte-generation devices is the ability to freely remap algorithmic qubits to ions due to the all-to-all connectivity of the devices. This hardware feature makes it simple to symmetrize the different errors associated with the locations of each ion. 

The algorithm for aggregation by non-linear filtering (outlined in \sfig~\ref{fig:em_dnl}) is given below:
\begin{enumerate}
    \item For each bit string obtain a sorted distribution of observed 
    frequencies of this bit string
    by each twirled variant (\sfig~\ref{fig:em_dnl}a). 
    \item Transpose the distribution of observed frequencies per variant to obtain the number of variants observing the given bit string with at least a specified frequency (\sfig~\ref{fig:em_dnl}b). The normalized area under that curve corresponds to the frequency of observing a given bit string if the data from the variants are directly merged without filtering. Note that the variants are considered individually regardless of how they were generated.
    \item Apply a threshold-based filtering function that discards bit strings not observed by the minimum number of variants. Higher thresholds allow for better mitigation of the biases introduced by the different impact of hardware noise across the variants, but are limited by the number of variants and shots taken (see \ref{app:MCMethod}). Since the observables are calculated on the lepton register, the quark register was ignored at this step.\label{enum:algorithm_step}
    \item Calculate the aggregated probability estimates as the normalized area under the curve after this filtering has been applied.
\end{enumerate}
\begin{figure}[h!]
\centering
  \includegraphics[width=0.75\linewidth]{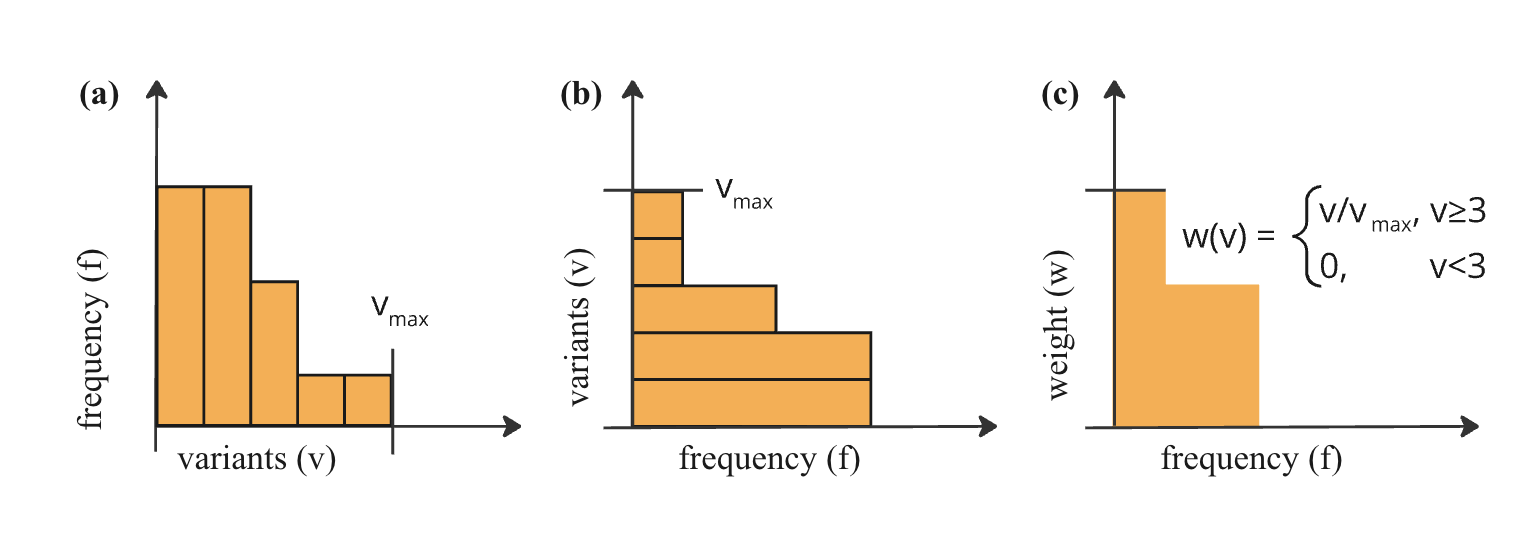}
  \caption{Aggregation with non-linear filtering. For each measured bit string, a distribution of frequencies per variant (a) is sorted from largest to smallest to be converted into a function of the number of individual variants simultaneously observing the given bit string at a given frequency (b). Normalized area under that curve leads to the aggregated results that correspond exactly to the aggregation by simple averaging if no filter is applied. (c) Filtering out any contributions observed by fewer than a certain minimum number of variants (3 on the picture) rules out accidental and biased counts.}
  \label{fig:em_dnl}
\end{figure}

This algorithm reduces the effect of bit strings that appear more frequently in only a few variants. This is an effective error mitigation strategy as all of the variants correspond to the same ideal quantum circuit, meaning the appearance of bit strings with higher frequencies in only a small subset of the variant histograms is likely due to noise. The threshold parameter introduced in Step~\ref{enum:algorithm_step} determines the size of this subset. In other words, the computed probability estimation depends more strongly on how common a bit string is among the variants than its frequency in a given histogram, especially when working with small numbers of shots.

Since the post-selection of bit strings in this method relies on the overlap between different variants, it follows that in the worst-case scenario, finding a non-zero overlap would require exponentially many measurements. In the middle-case scenario, this effect can be mitigated by leveraging the properties of observable instead of trying to refine the original measurement statistics.
For example, filtering can also be applied to binned subsets of bit strings to enhance the resolution of local observable if the shot count is low. 
The qubits not used for estimating that observable can be ignored by binning the outputs based on the measurements of the relevant qubits. This approach was used when working with the high depth circuit discussed in Results.
More symmetries can be used in larger scale simulations to increase the overlap between the variants. Binning can also be adjusted based on the targeted noise channels while others can be addressed with error detection.

\section{Monte Carlo Method for Determining the Variant Count}
\label{app:MCMethod}
\noindent 
When preparing to run the quantum simulation with circuit variants, we first need to determine how many variants to use and how many shots to take for each variant. A typical constraint on this optimization problem is the allocated runtime for the simulation on the quantum computer, and this is the constraint we used in making decisions for this project. Given a time budget, we estimate the number of shots that could be taken within that budget for a given number of variants based on the quantum gate times and related overheads. With this relationship between variants and shot count established, we then seek to minimize the expected absolute error in some output of a given combination. In this work we used the lepton number observable as the target output to optimize for.

Under the assumption that we do not know the ideal distribution of measured bit strings or the precise impact of hardware errors, we can approach this error estimate by marginalizing over both the ideal distribution and the distortion due to errors. This marginalization was performed with a Monte Carlo integration over prior probability distributions that describe this ignorance. We modeled these priors with symmetric Dirichlet distributions. That is, the probability vectors describing the output distributions were drawn from the probability density $f(\vec{p})$
\begin{equation}
    f(\vec{p}) = \frac{\Gamma(d\ \overline{\alpha})}{\Gamma(\overline{\alpha})^d}\prod_{i=1}^d p_i^{\overline{\alpha}-1}
\end{equation}
where $d$ is the dimension of the probability vector $\vec{p}$ and $\overline{\alpha}$ is the concentration parameter. (Note that $d$ can be much smaller than the full state-space of the quantum computer. An upper bound on the output dimension, for example, can be the number of shots taken.) To denote ignorance of the ideal probabilities we set $\overline{\alpha}=1$ so that this density describes a uniform distribution over all possible probability vectors.

We used a simplified model that is assumed to be close to global depolarizing noise to model the impact of noise. For global depolarizing noise, when an error occurs the measurement is of the maximum entropy state, meaning that the probability distribution over bit strings is uniform. Given that we expect to find notable biases in the output due to noise, we do not want to enforce global depolarizing noise exactly, so we instead again used the symmetric Dirichlet distribution and set $\overline{\alpha}=10$, which has the uniform probability vector as it's mode and but models error distributions that add some bias to global depolarizing noise. 

With these distributions and device error rates we can sample possible (mock) sets of histograms for a set of circuit variants. Each such set is a single Monte Carlo sample, and is drawn with the following steps:
\begin{enumerate}
    \item Draw a single mock ``ideal" probability vector as well as a set of ``noise" probability vectors.
    \item Draw from a binomial distribution (with the probability set by the device's error rate) how many shots are noisy for each variant.
    \item Construct a histogram for each variant by sampling the ideal distribution and variant-specific noise distribution a number of times determined in the previous step.
\end{enumerate}
For each Monte Carlo sample the absolute error (after the error mitigation was applied) is then recorded. These sample absolute errors are then averaged over, resulting in approximate Monte Carlo integration to compute the expected absolute error in an observable. Note that this expected absolute error models the influence of both shot noise and hardware errors.

By focusing on the lepton number observable and evaluating this expected absolute error for a wide set of possible numbers of variants, we settled on using $64$ variants with $150$ shots each for circuits run on Forte Enterprise. This choice was not made solely based on the expected absolute error calculation but also incorporated previous experiments with the DNL error mitigation method to select among the possible variant counts that were predicted to have a good performance.

We note that the the actual number of variants used was higher (see Results) as we ended up being able to allocate additional system time for this project beyond the original estimate.

\section{Quantum Circuits for State Preparation and Time Evolution}
\label{app:tevolCircs}
\noindent
In this supplementary note, we present the techniques used to construct the quantum circuits used in this work.
This includes circuits 
to
prepare $\vert \psi_{\text{init}} \rangle=|\psi_{\text{vac}}^{(\text{lep})}\rangle|\Delta^- \Delta^-\rangle$, and implement Trotterized time evolution.
The circuits were
further optimized when transpiling to IonQ's native gate set using the IonQ circuit compiler.

\subsection{State Preparation}
\noindent
The initial two-baryon  state is
$\vert \psi_{\text{init}} \rangle=|\psi_{\text{vac}}^{(\text{lep})}\rangle|\Delta^- \Delta^-\rangle$, where $|\psi_{\text{vac}}^{(\text{lep})}\rangle$ is the ground state of 
$\hat{H}_{\text{leptons}}$ 
given in \seq~\eqref{eq:BetaHamL2_lepkin}.
Circuits for preparing the lepton vacuum on $L=2$ spatial sites with PBCs are shown in \sfig~\ref{fig:LeptonVacInit}. 
This circuit uses layers of Givens rotations followed by fermionic SWAPs to exactly prepare the free fermion ground state~\cite{Jiang:2017pyp,Wecker_2015,Kivlichan_2018}.
The construction of this circuit follows several principles.
First, the relative fermionic phases between electrons and neutrinos are pushed to the end of the circuit.
This allows the circuits that initialize the electron and neutrino vacuums to be determined separately, since $|\psi_{\text{vac}}^{(\text{lep})}\rangle=|\psi_{\text{vac}}^{(e)}\rangle |\psi_{\text{vac}}^{(\nu)}\rangle$ if the fermionic statistics are ignored. 
The electron and neutrino ground states are real, translationally invariant and have equal fermion and anti-fermion occupation.
The circuit in \sfig~\ref{fig:LeptonVacInit}b) is able to reach all such states, and it does so with two components.
The first component applies an $R_Y$ rotation on the first qubit of each spatial lattice site followed by $\hat{X}$ on the second qubit controlled on the first qubit being in state $\ket{0}$. 
This ensures that the initial state is real and has equal fermion and anti-fermion occupations.
The second component produces all possible lepton-number conserving, real two-qubit transformations that the first layer does not account for. 
This is done using a pair of $(\hat{X}\hat{Y} - \hat{Y}\hat{X})$ rotations.
The $(\hat{X}\hat{Y} - \hat{Y}\hat{X})$ rotation that wraps around the lattice has a relative minus sign due to the minus sign in the JW kinetic term.
The resulting circuit exactly prepares the ground state and is parameterized by two angles $\{\theta_0,\theta_1\}$ that depend on the value of the fermion mass.
For the masses chosen in \seq~\eqref{eq:params}, the variational parameters are determined by minimizing the energy of the one-flavor lepton Hamiltonian, and are
\begin{equation}
    \{\theta_0, \theta_1 \}_{m_e = 0.1} \ = \ \{ 0.7356, -1.2030\} \ \ \ \, \ \ \ \ \{\theta_0, \theta_1 \}_{m_{\nu} = 1.5} \ = \ \{ 0.2940, -1.4238\} \ .
    \label{eq:varParams}
\end{equation}
Note that the parameters do not depend on the Majorana mass because $|\psi_{\text{vac}}^{(\nu)}\rangle$ is the same for the $m_\nu=1.5$, $m_M=0$ and $m_\nu=1.5$, $m_M=1.7$ used in this work.
The fermionic statistics are included at the end of the circuit using the $CZ$ network in \sfig~\ref{fig:LeptonVacInit}a).
\begin{figure}
    \centering
    \includegraphics[width=0.8\textwidth]{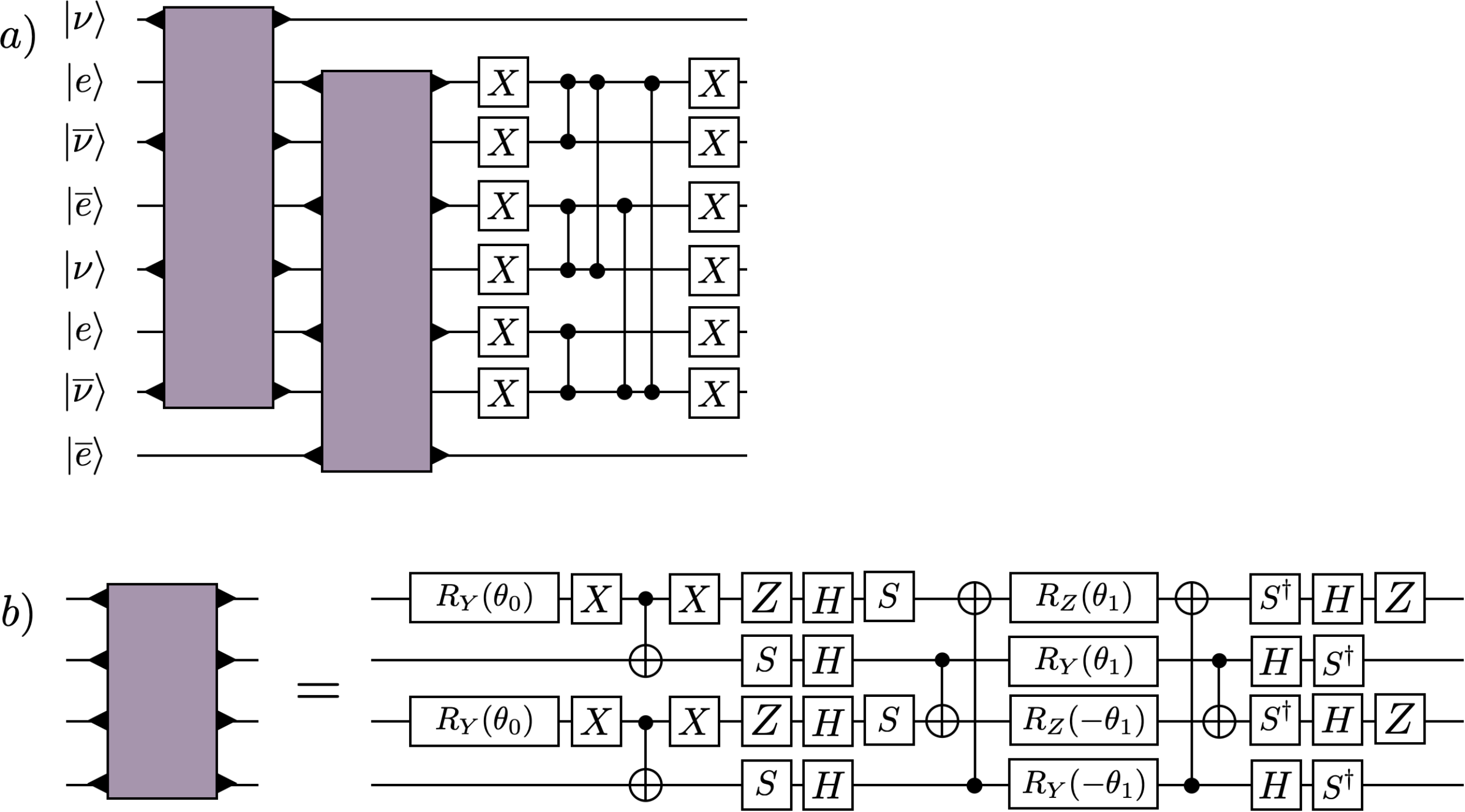}
    \caption{The circuits for preparing the two-flavor lepton vacuum on $L=2$. 
    The circuit in a) prepares the two-flavor vacuum using the one-flavor vacuum preparation circuit block defined in b).
    The parameters $\{\theta_0,\theta_1\}$ are determined by variationally minimizing the one-flavor lepton Hamiltonian, and are given in \seq~\eqref{eq:varParams} for $m_e=0.1$ and $m_{\nu}=1.5$.
    The barbells denote  $CZ$ gates and the \rotatebox[origin=c]{-90}{$\blacktriangle$}-symbols mark which qubits the circuit blocks are acting on.}
    \label{fig:LeptonVacInit}
\end{figure}

The initial state in the QCD sector is $|\Delta^- \Delta^-\rangle$,
which,
as mentioned in Methods~A, factorizes between the up- and down-quark sectors,\footnote{There are 
extra minus signs in the $u$-quark vacuum due to $u$-quarks hopping over three $d$-quarks.}
\begin{align}
\vert \Delta^- \Delta^-\rangle \  = \ \vert d_r d_g d_b\, d_r d_g d_b\rangle \vert \psi_{\text{ vac}}^{(u)}\rangle \ .
    \label{eq:L2B2}
\end{align}
The $\vert d_r d_g d_b\, d_r d_g d_b\rangle$ state is trivial to prepare, and circuits for approximately preparing $\vert \psi_{\text{ vac}}^{(u)}\rangle$ can be determined with the SC-ADAPT-VQE algorithm~\cite{Farrell:2023fgd,Farrell:2024fit}.
SC-ADAPT-VQE determines low-depth scalable circuits that efficiently prepare states with localized correlations using symmetries and hierarchies in length scales to define a variational circuit ansatz.
These circuits are optimized to minimize the energy using ADAPT-VQE~\cite{Grimsley_2019} running on a classical computer.
Here, SC-ADAPT-VQE is used to determine circuits that prepare $\vert \psi_{\text{ vac}}^{(u)}\rangle$.
The QCD Hamiltonian describing a single quark flavor is similar to \seq~\eqref{eq:HKSPBCv2},
\begin{align}
    \hat{H} =& \ \sum_{c=0}^2 \left \{\frac{1}{2}\sum_{n=0}^{2L-1}m_u\left [(-1)^n \hat{Z}_{3n+c}+\hat{I}\right ] 
    \ + \ 
    \frac{1}{2}\sum_{n=0}^{2L-2}\left [\hat{\sigma}^+_{3n+c}\hat{Z}^2\hat{\sigma}^-_{3n+3+c}+{\rm h.c.}\right ] 
    +\frac{1}{2}(-1)^{L+1}
    \left[\hat{\sigma}^+_{6L-3+c}\hat{Z}^2\hat{\sigma}^-_{c}+{\rm h.c.}\right]  \right \} 
    \nonumber \\[4pt]
    &-\frac{g^2}{2}\sum_{n=0}^{2L-1}\sum_{a=1}^{8}\sum_{s=1}^L\left (s-\frac{s^2}{2L} \right )\left ( 1-\frac{\delta_{s,L}}{2}\right )\hat{Q}_n^{(a)}\hat{Q}_{n+s}^{(a)} \ ,
\end{align}
where 
\begin{align}
    \sum_{a=1}^8 \hat{Q}_{n}^{(a)} \, \hat{Q}_{m}^{(a)}  =   \frac{1}{4}\bigg [ & \ 2\big (\hat{\sigma}^+_{3n}\hat{\sigma}^-_{3n+1}\hat{\sigma}^-_{3m}\hat{\sigma}^+_{3m+1} + \hat{\sigma}^+_{3n}\hat{Z}_{3n+1}\hat{\sigma}^-_{3n+2}\hat{\sigma}^-_{3m}\hat{Z}_{3m+1}\hat{\sigma}^+_{3m+2} 
    + \hat{\sigma}^+_{3n+1}\hat{\sigma}^-_{3n+2}\hat{\sigma}^-_{3m+1}\hat{\sigma}^+_{3m+2} + { \rm h.c.}\big ) \nonumber \\
    &+ \frac{1}{6}\sum_{c=0}^{2} \sum_{c'=0}^2( 3 \delta_{c c'} - 1 ) \hat{Z}_{3n+c}\hat{Z}_{3m+c'} \bigg ] \ .
    \label{eq:QnfQmfp_1f}
\end{align}

In ADAPT-VQE, the unitaries that prepare the vacuum are constructed from a pool of Hermitian operators $\{ {\hat O} \}$.
For vacuum preparation, the operators in the pool
respect translational invariance, global $SU(3)$, baryon number, parity, charge conjugation and time reversal.
The easiest way to construct operators that conserve these symmetries is by using terms in the Hamiltonian.
However, terms in the Hamiltonian are real and their unitary evolution $e^{i \theta \hat{O}}$ is not optimal for preparing the vacuum, which is a real wavefunction.
A good starting point is the commutator of terms in the Hamiltonian 
(multiplied by ``$i$'' to make them hermitian).\footnote{For preparing the vacuum of the Schwinger model, 
an operator pool generated from the algebra of the non-interacting Hamiltonian was found to be effective~\cite{Farrell:2023fgd,Farrell:2024fit,Farrell:2024mgu}.}
This can be generalized to $SU(3)$, and a suitable pool of operators is
\begin{align}
\{\hat{O} \} \ &= \ \{\hat{O}_{mh}(s)\} \ , \nonumber \\[4pt]
\hat{O}_{mh}(s) &= i \left [\hat{\Theta}_m, \hat{\Theta}_h(s) \right ]\ , \nonumber \\[4pt]
\hat{\Theta}_m &= \frac{1}{2}\sum_{n=0}^{N-1}\sum_{c=0}^2 (-1)^n \hat{Z}_{3n+c}\ , \nonumber \\[4pt]
\hat{\Theta}_h(s) &= \sum_{n=0}^{N-1}\sum_{c=0}^2 v(n,s)(1-\frac{1}{2}\delta_{s,L})\left [\hat{\sigma}^+_{3n+c} \hat{Z}^{3s-1}\hat{\sigma}^-_{3(n+s)+c} + {\rm h.c.}\right ]\ ,
\label{eq:QCDpool}
\end{align}
where $\hat{\Theta}_h(s)$ is an operator that hops $s$ staggered sites.
The phase $v(n,s) = (\pm 1)^{L+1}$ with $(+)$ if $3(n+s) < 3N-1$ and $(-)$ if $3(n+s) > 3N-1$ comes from the minus sign in the JW mapped kinetic term.
The range of $s$ is $s\in \{1,3,\ldots,L\}$, and only $s$ odd operators are generated as a consequence of charge conjugation symmetry.
This pool of operators creates mesonic (quark-antiquark) 
excitations, and are found to provide rapid
convergence to the vacuum for the parameter regime considered.
The initial state for the SC-ADAPT-VQE preparation of the vacuum 
is
the strong-coupling vacuum $\vert \Omega_0 \rangle = \vert 000\,111\rangle^{\otimes L}$.
Surprisingly, as it is not the case with OBCs, the $\hat{O}_{mh}(s)$ with different (odd) $s$ commute with each other, i.e., $[\hat{O}_{mh}(s), \hat{O}_{mh}(s')]=0$.
Therefore, in the absence of Trotter errors, SC-ADAPT-VQE converges once all values of $s$ are explored ($\lceil L/2 \rceil$ steps).
Trotter errors will break this and SC-ADAPT-VQE may
require more steps to converge.

Two quantities are used to 
quantify the quality of the SC-ADAPT-VQE prepared vacuum $|\psi_{\text{ans}}\rangle$.
The first is the deviation in the energy of the ansatz state $E_{\text{ans}}$ compared to the true vacuum energy $E_{\text{vac}}$,
\begin{align}
\delta E = \frac{E_{\text{vac}} - E_{\text{ans}}}{E_{\text{vac}}}
\ ,
\end{align}
and 
the second is the infidelity density,
\begin{align}
{\cal I}_L =\frac{1}{L}\left (1 - \vert \langle \psi_{\text{vac}} \vert  \psi_{\text{ans}} \rangle \vert^2  \right ) 
\ .
\end{align}
Results for the energy and infidelity density for $m=g=1$ (with a correlation length of $\xi \propto m_{\text{hadron}}^{-1} = 0.34$ staggered sites, where $m_{\text{hadron}}$ is the lightest hadron mass) are given in \stab~\ref{tab:AdaptEF}. 
The sequence of operators and their corresponding variational parameters are given in \stab~\ref{tab:AdaptAng}.
As expected, SC-ADAPT-VQE initially builds out short-range correlations (small $s$), which have the largest variational parameters.
The variational parameters are converging with the expected $e^{-L/\xi}$ scaling and could be robustly extrapolated to prepare the vacuum for large $L$.
As mentioned above, the SC-ADAPT-VQE algorithm with this pool would converge after $\lceil L/2 \rceil$ steps in the absence of Trotterization errors. 
We find that the Trotter errors are small, and our implementation of SC-ADAPT-VQE also converges after $\lceil L/2 \rceil$ steps using a convergence criteria of $\vert \langle \psi_{\text{ans}}\vert  [\hat{H},\hat{O} ] \vert \psi_{\text{ans}}\rangle \vert \leq 10^{-5}$.
Despite rapid
convergence, the ultimate wavefunction fidelity exceeds that required for near-term quantum simulations.

The circuits that implement the unitary evolution of the operators in \seq~\eqref{eq:QCDpool} are similar to those used for preparing the Schwinger model vacuum in Ref.~\cite{Farrell:2023fgd}.
Those circuits utilized the ``X"-circuit design of Ref.~\cite{Algaba:2023enr} to minimize circuit depth on a device with nearest-neighbor connectivity.
However, for devices with all-to-all connectivity like IonQ Forte, the technique using $CZ$s that is described in \ref{sec:tevol} below is superior.
An example of the circuit that prepares the SC-ADAPT-VQE vacuum for one quark flavor with $L=2$ 
is shown in \sfig~\ref{fig:VacPrepCircs}a).
This circuit has a CNOT depth of 8 (compared to 11 with nearest-neighbor connectivity).
\begin{figure}
    \centering
    \includegraphics[width=\textwidth]{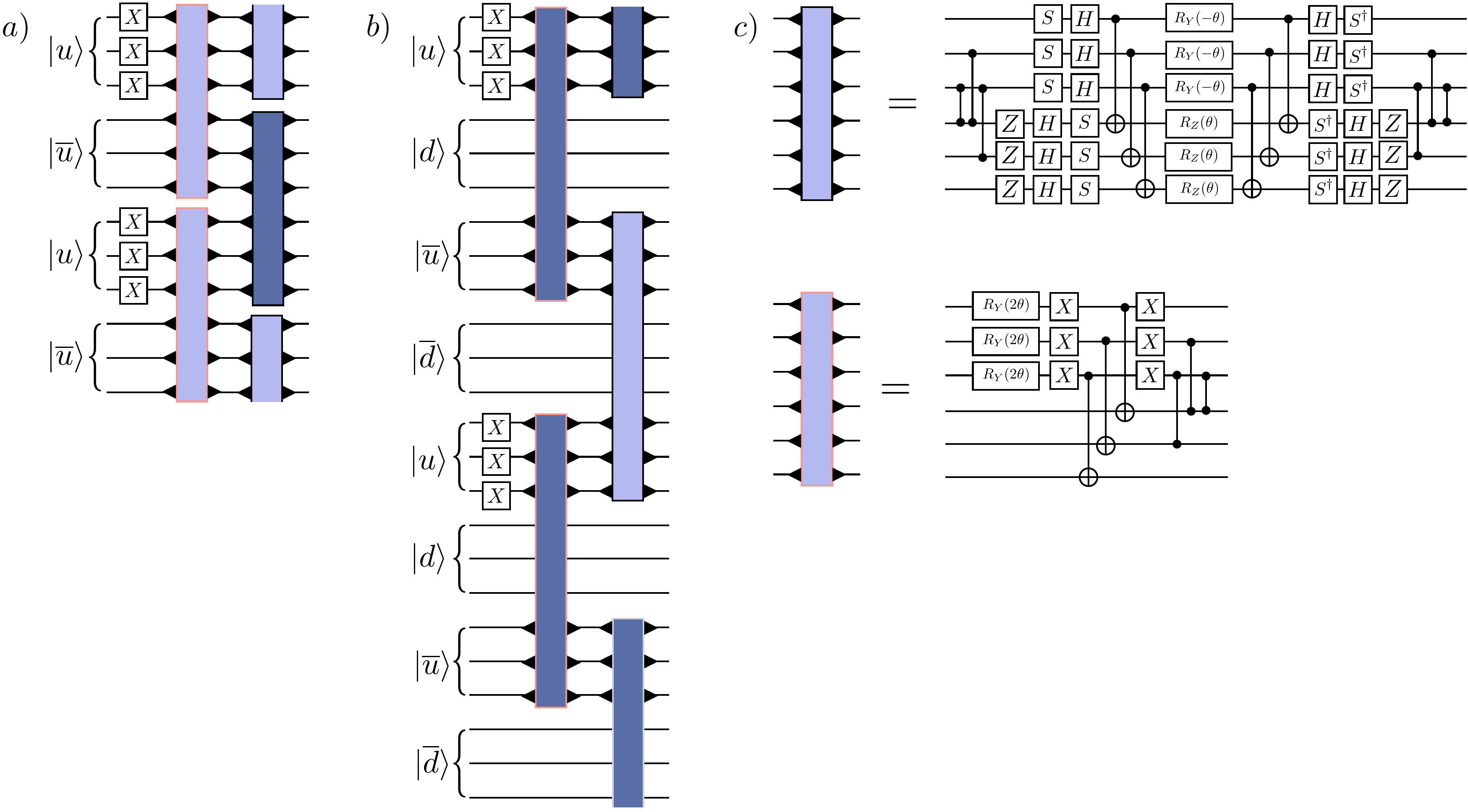}
    \caption{The SC-ADAPT-VQE circuits used for state preparation in $1+1$D QCD on a $L=2$ lattice. 
    The circuit in a) prepares $|\psi_{\text{vac}}^{(u)}\rangle$, and the circuit in b) prepares $|\Delta^- \Delta^-\rangle$.
    The circuits in c) implement $e^{-i\frac{\theta}{2}\left (\hat{X}\hat{Z}\hat{Z}\hat{Y}-\hat{Y}\hat{Z}\hat{Z}\hat{X} \right )}$ and the pink-outlined circuit blocks are a simplification of the black outlined circuit blocks when acting on the strong-coupling vacuum initial state $|\Omega_0\rangle$.
    The dark purple blocks are the same as the light purple ones except with $\theta \to - \theta$.
    The value of $\theta$ for $m=g=1$ and $L=2$ is given in \stab~\ref{tab:AdaptAng}.
    The barbells denote  $CZ$ gates and the \rotatebox[origin=c]{-90}{$\blacktriangle$}-symbols mark which qubits the circuit blocks are acting on.
    }
    \label{fig:VacPrepCircs}
\end{figure}
\begin{table}[!t]
\renewcommand{\arraystretch}{1.4}
\begin{tabularx}{\textwidth}{c | Y Y Y Y Y Y Y Y }
 \hline \hline
 \multicolumn{1}{c}{} & \multicolumn{4}{c}{$\delta E $ } & \multicolumn{4}{c}{${\cal I}_L $} \\
  \cmidrule(lr){2-5} \cmidrule(lr){6-9}
 \diagbox[height=23pt]{$L$}{\text{step}} & 0 & 1 & 2 & 3 & 0 & 1 & 2 & 3 \\
 \hline 
 2 & 0.3462 & 0.0002 &  &  &  0.1492 & 0.00004 &  & \\
 \hline
 3 & 0.3498 & 0.0014 & 0.0002 &  & 0.1431 & 0.0006 & 0.0001 & \\
 \hline
 4 & 0.3471 & 0.0007 & 0.0002 &  & 0.1293 & 0.0003 & 0.0001 & \\
 \hline
 5 & 0.3459 & 0.0007 & 0.00023 & 0.00022 & 0.1187 & 0.0003 & 0.000061 & 0.000064\\
 \hline \hline
\end{tabularx}
\caption{The convergence of up to three steps of the SC-ADAPT-VQE algorithm for $L=2-5$ and $m=g=1$. 
For $L=2$, SC-ADAPT-VQE has converged after one step, for $L=3$ and $L=4$, SC-ADAPT-VQE has converged after two steps, and for $L=5$ it has converged after three steps. 
The deviation in the energy $\delta E$ and the infidelity density ${\cal I}_L$ are defined in the text.
Step zero represents the initial state $\vert \psi_{\text{ans}} \rangle = \vert \Omega_0 \rangle$.}
 \label{tab:AdaptEF}
\end{table}
\begin{table}[!t]
\renewcommand{\arraystretch}{1.4}
\begin{tabularx}{0.5\textwidth}{c | Y Y Y}
 \hline \hline
 \diagbox[height=23pt]{$L$}{$\theta_i$} & $\hat{O}_{mh}(1)$ & $\hat{O}_{mh}(3)$ & $\hat{O}_{mh}(5)$  \\
 \hline 
 2 & 0.1705 &  &  \\
 \hline
 3 & 0.1755 & $-0.0137$ &  \\
 \hline
 4 & 0.1732 & $-0.0058$ &  \\
 \hline
 5 & 0.1723 & $-0.0059$ & 0.0008 \\
 \hline \hline
\end{tabularx}
\caption{
The order of the operators $\hat{O}$ and variational parameters $\theta_i$ used to prepare the SC-ADAPT-VQE vacuum in 1+1D QCD with one flavor of quark.
Shown are results for $L=2-5$ and $m=g=1$. }
 \label{tab:AdaptAng}
\end{table}

The circuits that prepare $|\psi_{\text{vac}}^{(u)}\rangle$ on $L=2$ can immediately be used to prepare $|\Delta^- \Delta^-\rangle$.
The down quarks sites
are fully occupied, 
which causes the signs of the variational parameters in the $|\psi_{\text{vac}}^{(u)}\rangle$ preparation circuit to be negated. 
The circuits used to prepare the $|\Delta^- \Delta^-\rangle$ state are shown in \sfig~\ref{fig:VacPrepCircs}b).
The initial state for our quantum simulations 
$\vert\psi_{\text{init}} \rangle=|\psi_{\text{vac}}^{(u)}\rangle|\Delta^- \Delta^-\rangle$ 
is prepared by applying the circuit in \sfig~\ref{fig:VacPrepCircs}b) to qubits 
$q_0,q_1,\ldots,q_{23}$ and the lepton vacuum initialization circuit in 
\sfig~\ref{fig:LeptonVacInit}a) to qubits $q_{24},q_{25},\ldots,q_{31}$ (see layout in Fig.~4 in the main text).

\subsection{Time Evolution}
\label{sec:tevol}
\noindent
In our previous work, circuits for digitizing the time-evolution operator were developed, making use of an ancilla to store partial parities of the JW strings~\cite{Farrell:2022wyt,Farrell:2022vyh}.
An improved method for creating these circuits without an ancilla will be presented here~\cite{Chernyshev:2025jyw, Kivlichan_2018, CerveraLierta2018exactisingmodel}.
These circuits are arranged to maximize parallelization, hence reducing the circuit depth.
The idea is to design the circuits without the JW $\hat{Z}$ strings, and then add them in at the end.
Important circuit identities are: 
\begin{align}
 CZ (\hat{Y} \otimes \hat{I} ) CZ &= \hat{Y} \otimes \hat{Z} \ , \nonumber \\
 CZ (\hat{X} \otimes \hat{I} ) CZ &= \hat{X} \otimes \hat{Z} \ , \nonumber \\
 (CZ)^2 &= \hat{I} \ , \nonumber \\
 CZ_{i,i+1} &= CZ_{i+1,i} \ .
\end{align}
Making use of these identities allows the JW $\hat{Z}$s to be put in by sandwiching the circuit with a sequence of $CZ$s.
An illustrative example of this method was given in Fig.~5 in the main text.
\begin{figure}
    \centering
    \includegraphics[width=\textwidth]{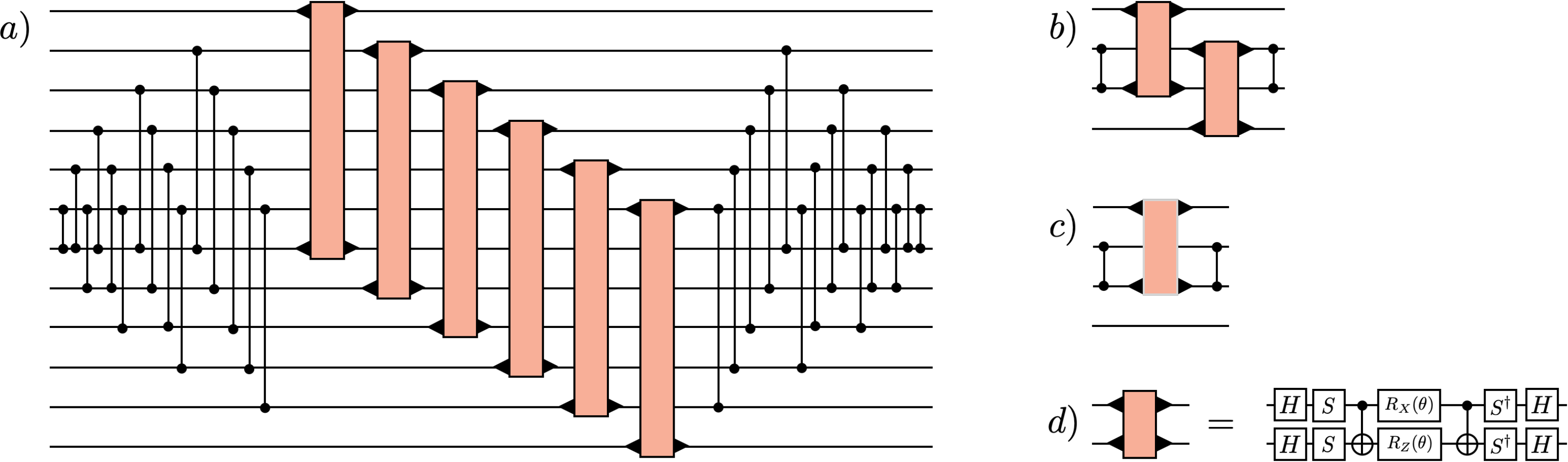}
    \caption{Quantum circuits that implement the unitary evolution from the kinetic and Majorana mass terms in the Hamiltonian.
    The barbells denote  $CZ$ gates and the 
    \rotatebox[origin=c]{-90}{$\blacktriangle$}-symbols 
    on the orange blocks mark the qubits that are acted on.
    a) A quantum circuit that implements the unitary evolution of the quark kinetic term in \seq~\eqref{eq:BetaHamL2_quarkkin} across one spatial site.
    b) A quantum circuit that implements the unitary evolution of the lepton kinetic term in \seq~\eqref{eq:BetaHamL2_lepkin} across one spatial site.
    c) A quantum circuit that implements the unitary evolution of the neutrino Majorana mass term in \seq~\eqref{eq:BetaHamL2_lepkin} across one spatial site.
    The light-gray border on the circuit blocks denotes that the the rotation in d) is $R_X(-\theta)$.
    d) The definition of the light orange circuit block that implements $e^{-i\theta( \hat{X}\hat{X}+\hat{Y}\hat{Y})/2}$.}
    \label{fig:kin_CZ_circ}
\end{figure}
The extension of this method to generate a circuit that implements the quark kinetic term in \seq~\eqref{eq:BetaHamL2_quarkkin} is shown in \sfig~\ref{fig:kin_CZ_circ}b), that implements the lepton kinetic term in \seq~\eqref{eq:BetaHamL2_lepkin} is shown in \sfig~\ref{fig:kin_CZ_circ}c) and that implements the Majorana mass term in \seq~\eqref{eq:BetaHamL2_lepkin} is shown in \sfig~\ref{fig:kin_CZ_circ}d).
Additionally, combining this method with the circuits from Refs.~\cite{Farrell:2022wyt,Farrell:2022vyh,Stetina:2020abi}, allows the weak interactions in \seq~\eqref{eq:BetaHamL2_beta} to be implemented. 
This is shown in \sfig~\ref{fig:BetaCirc} and significantly reduces the circuit depth relative to the circuits in Ref.~\cite{Farrell:2022vyh}. For the strong-interaction terms in \seq~\eqref{eq:BetaHamL2_glue}, we use the circuits from Ref.~\cite{Farrell:2022wyt}.

\begin{figure}[t!]
    \centering
    \includegraphics[width=\textwidth]{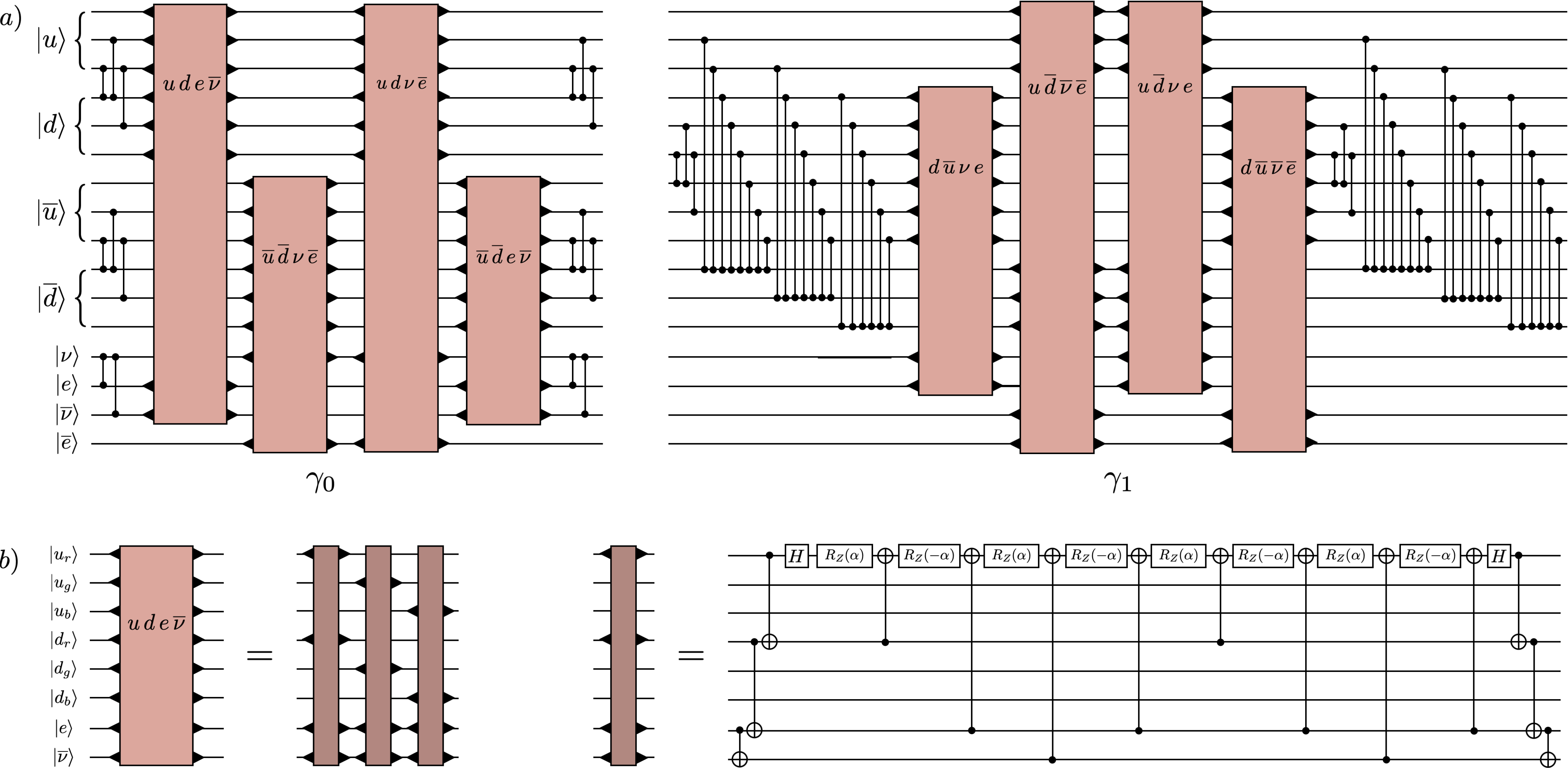}
    \caption{a) The quantum circuits used to implement the unitary evolution from the $\beta$-decay interaction in \seq~\eqref{eq:BetaHamL2_beta} across one spatial site.
   The barbells denote  $CZ$ gates and the \rotatebox[origin=c]{-90}{$\blacktriangle$}-symbols on the light-brown box mark the qubits that are acted on.
    The circuits have been split into the $\gamma_0$ terms (first and second lines of $H_{\beta}$ in \seq~\eqref{eq:BetaHamL2_beta}), and the $\gamma_1$ terms (third and fourth lines of $H_{\beta}$ in \seq~\eqref{eq:BetaHamL2_beta}).
    The circuit blocks are labeled by the fermion sites that they act on and have been ordered to allow for maximum parallelizability.
    b) The definition of the light-brown circuit block using the GHZ diagonalization circuits (dark-brown circuit block) that implements $e^{-4i\alpha (\hat{\sigma}^+_i\hat{\sigma}^-_j\hat{\sigma}^+_k\hat{\sigma}^-_l +\, {\rm h.c.} )}$. 
    This example is for the $ude\overline{\nu}$ term, and the rotation angle is $\alpha = \sqrt{2} G t/8$.
    The other terms implement permutations of this circuit, that depend on the ordering of $\{i,j,k,l\}$ in $(\hat{\sigma}^+_i\hat{\sigma}^-_j\hat{\sigma}^+_k\hat{\sigma}^-_l+{\rm h.c.})$.}
    \label{fig:BetaCirc}
\end{figure}

\endgroup

\end{document}